\documentclass[aps,notitlepage,nofootinbib,longbibliography,
twocolumn,
superscriptaddress,10pt]{revtex4-1}
\usepackage[colorlinks=true,citecolor=blue,linkcolor=blue]{hyperref}
\usepackage[normalem]{ulem}
\usepackage[utf8]{inputenc}
\usepackage[german, english]{babel}
\usepackage[dvipsnames]{xcolor}
\usepackage{amsmath,amsfonts, amssymb, amsthm, dsfont}
\usepackage{bm}
\usepackage{graphicx}
\usepackage{tikz}
\usepackage{physics}
\usepackage{xspace}
\usepackage[export]{adjustbox}

\usepackage{enumerate}
\usepackage{subcaption}
\usepackage{comment}
\usepackage{makecell}
\usepackage[justification=raggedright,format=plain]{caption}

\newcommand{\sgn}{\operatorname{sgn}}

\newcommand{\Z}{\mathbb{Z}}

\newcommand{\be}{\begin{equation}}
\newcommand{\ee}{\end{equation}}
\newcommand{\ba}{\begin{aligned}}
\newcommand{\ea}{\end{aligned}}

\newcommand{\syan}[1]{{\color{cyan} 
\textbf{SWY: #1}}}

\graphicspath{{Figures/}}

\begin{document}

\title{
Dissipative Dynamical Phase Transition as a Complex Ising Model}
\author{Stephen W. Yan}
\affiliation{Department of Physics, University of California,
Santa Barbara, CA 93106, USA}
\author{Diego Barberena}
\affiliation{JILA, Department of Physics, University of Colorado,  Boulder, CO 80309, USA}
\affiliation{Center for Theory of Quantum Matter, University of Colorado, Boulder, CO 80309, USA}
\affiliation{T.C.M. Group, Cavendish Laboratory, University of Cambridge, J.J. Thomson Avenue, Cambridge CB3 0HE, UK}
\author{Matthew P. A. Fisher}
\affiliation{Department of Physics, University of California,
Santa Barbara, CA 93106, USA}
\author{Sagar Vijay}
\affiliation{Department of Physics, University of California,
Santa Barbara, CA 93106, USA}
\date{\today}

\begin{abstract}
We investigate a quantum dynamical phase transition induced by the competition between local unitary evolution and dissipation in a qubit chain with a strong, on-site $\Z_2$ symmetry.  While the steady-state of this evolution is always maximally-mixed, we show that the dynamical behavior of certain non-local observables on the approach to this steady-state is dictated by a quantum Ising model with a \emph{complex} transverse-field (cTFIM). 
A crucial feature of our mapping to the cTFIM is an extensive number of local, weak, $\mathbb{Z}_2$ symmetries possessed by the dynamics, allowing us to investigate these observables analytically and uncover a dynamical phase transition as the relative rate of unitary evolution and dissipation is tuned.
  We show that the weak-dissipation limit corresponds to a cTFIM with a large magnitude of the imaginary transverse-field, for which the many-body ``ground-state" (with smallest real eigenvalue) is gapless, exhibiting quasi-long-range correlations of the local magnetization with a continuously-varying exponent.  Correspondingly, the dynamics of the non-local observables show oscillatory behavior with an amplitude decaying exponentially in time.  
The strong-dissipation limit corresponds to a gapped ferromagnetic phase of the cTFIM, and non-local observables show exponential decay on the approach to equilibrium.
This transition in (1+1)-dimensions has a peculiar, ``two-sided" nature appearing as either first- or second-order depending on the phase from which the transition is approached, an analytic result which is 
corroborated by numerical studies.  In higher dimensions, we present a field-theoretic understanding of the first-order nature of this transition, when approaching from the ferromagnetic phase of the cTFIM, though the nature of the phase with large imaginary transverse-field remains to be understood.
\end{abstract}

\maketitle
\tableofcontents

\section{Introduction}
The recent advent of Noisy Intermediate-Scale Quantum (NISQ) Devices \cite{preskill2018quantum}
has allowed for the experimental realization and characterization of new quantum phases that 
are difficult to observe or lack a counterpart in more traditional systems.
Experimentally, NISQ devices allow for highly controllable dynamics, including the ability to directly monitor the microscopic degrees of freedom and apply conditional feedback in order to mitigate decoherence or to steer towards a desired steady state \cite{altman2021quantum}. 

There have been many recent investigations into the consequences of the extended class of quantum many-body dynamics allowed by the NISQ toolbox.  Key features enabled by these ingredients, such as novel  entanglement structures~\cite{Skinner_MIPT,Li_MIPT,Amos_MIPT,li2023entanglement,gullans2020dynamical,chen2024symmetry,wang2024anomaly,zhang2024quantum,kim2024persistent}, phases of open quantum matter~\cite{sang2024stability,Ma_average_SPT,coser2019classification,sang2024mixed,de2022symmetry,lavasani2024stability,rakovszky2024defining,ellison2024towards,wang2023intrinsic,zhang2024strong,bao2023mixed,ma2023topological}, and quantum collective phenomena ~\cite{tantivasadakarn2023hierarchy, tantivasadakarn2024long,lu2023mixed,hastings2021dynamically,sala2024quantum,lee2023quantum,zhu2023nishimori,garratt2023measurements}, are being explored.
In this paper, we present a complementary approach where the mixed dynamics of a $(1+1)$-d qubit chain under both coherent unitary evolution and dissipation induced by measurement of the system by an environment can stabilize a featureless, maximally-mixed state at late times, but the nature of the system's approach to that state can be non-trivial.
The behavior of the system at intermediate times is probed by a linear, but non-local observable of the density matrix, which in the thermodynamic limit undergoes a sharp transition between ``underdamped oscillations" and ``overdamped decay" as the relative strength of the unitary versus dissipative evolution is tuned.

The dynamics that we investigate can, in principle, be implemented through discrete-time dynamics (unitary evolution and measurements) on a quantum simulator.  We analytically study this evolution in a continuous-time limit, where we find that the physics is governed by an effective non-Hermitian Hamiltonian (NHH). This NHH provides a precise description of the superoperator governing the evolution of the mixed quantum many-body state; we emphasize that this is in contrast to NHHs which often appear in the literature as an approximation to the full Lindbladian evolution when neglecting quantum jumps, and which are accessible through post-selection~\cite{Matsumoto_YL_exp,Lee_YL_exp,Gao_YL_exp,Naghiloo_YL_exp}.

A crucial feature of the dynamics that we investigate is the presence of a weak symmetry, which leads to a ``fragmentation" of the doubled Hilbert space, within which the density matrix of the system evolves \cite{essler2020integrability,
PhysRevResearch.5.043239,PhysRevB.109.054311}.  This property guarantees that certain observables only evolve within a subspace of the full operator Hilbert space under the dissipative evolution. Furthermore, the time-evolution of distinct, non-local observables can be governed by distinct NHHs. 
The main result of this paper is the study of a particular restricted class of observables, whose time-evolution can be exactly solved analytically. We find that the expectation value of our observable at time $t$ has the interpretation as a ``return amplitude" of the initial state following a quantum quench into imaginary-time evolution for time $t$ under the effective NHH. Such quantities have been shown to exhibit dynamical quantum phase transitions~\cite{Heyl_DQPT_review,Heyl_DQPT_TFIM,Jurcevic_DQPT_observation,Flaschner_DQPT_cold_atom,Budich_DQPT_topological,Sharma_DQPT_topological}, displaying non-analytic behavior at finite times and system sizes.

The time-dependence of our observables are characterized by an oscillation frequency $\omega$ and a exponential decay rate $\Gamma$.
We find that the oscillatory behavior $\omega$ is sensitive to short-distance details such as the parity of the system size $L$ and lacks a well-defined thermodynamic limit.
The non-analyticity in $\omega$ is determined by an exceptional point~\cite{Kato_Perturbation_Theory_Lin_Operators,Heiss_Physics_Exceptional_Points,Heiss_Level_Crossing,Heiss_Phases_Of_Wavefunctions,Moiseyev_NH_QM} which is present even in a finite-size system.
In contrast, we show that the second derivative of the decay rate is singular only in the thermodynamic limit and is a many-body effect.

An alternative perspective is to study the ground state\footnote{The relevant ``ground-state" is the right-eigenstate of the NHH with the smallest real part of its eigenvalue.} properties of the effective NHH, which is the Ising model with complex transverse field (cTFIM).  
The ``overdamped" phase of the dynamics is identified with a gapped ferromagnetic phase exhibiting long-range order, but with a magnetization that drops discontinuously to zero at the transition.
On the other hand, the ``underdamped" phase is identified with a gapless, power-law-correlated phase with continuously varying exponents.
We show that these correlations in the right-eigenstate of the NHH can be measured via the expectation values of the same set of non-local observables but subject to a slightly modified time evolution.

The rest of the paper will be organized as follows.
In Section~\ref{sec:zero_dim}, we will review the problem of a single qubit under mixed unitary and dissipative dynamics, and introduce the basic physics of the ``underdamped/overdamped" transition.
In~\ref{sec:one_dim_model_dynamics}, we introduce a 
$1$-d generalization of the model 
and demonstrate how observables linear~\ref{sec:one_dim_linear_observable} and non-linear~\ref{sec:one_dim_non_linear} in the density matrix probe quantities governed by the cTFIM.
We describe the qualitative behavior of the two phases in~\ref{sec:analytic_qualitative_behavior}, study the singular behavior of the late-time decay rate and oscillation frequency near the transition in~\ref{sec:properties_one_dim_model}, followed by numerical studies of our analytic solution at finite times in~\ref{sec:one_dim_numerics} and studies of the ground state spin-spin correlation functions in~\ref{sec:spin_spin_correlations_1d}.
In Section~\ref{sec:field_theory}, we introduce a field theoretic approach to
the general problem in higher dimensions, which captures qualitative features of the transition approached from the overdamped phase.
Finally, we conclude in~\ref{sec:conclusion}.
\section{Warmup: Single-Qubit Dynamics}\label{sec:zero_dim}
In this section, we review a canonical example of a $2$ x $2$ non-Hermitian Hamiltonian (NHH) demonstrating the physics of an exceptional point transition characterized by spontaneous $\mathcal{P}\mathcal{T}$-symmetry breaking (see for example,~\cite{Naghiloo_YL_exp, joglekar2018passive}).
We frame this discussion with language that will be useful when we introduce the one-dimensional model (Section~\ref{sec:one_dim}).
\subsection{Model Dynamics}\label{sec:zero_dim_model_dyn}
We consider a single spin under discrete-time evolution
with each time step $\Phi_{p, \theta, \delta t}$ 
consisting of unitary rotation $U_{\theta, \delta t}$ followed by decoherence channel $M_{\mu; p, \delta t}$
\begin{equation}\label{eqn:zero_dim_single_time_step}
    \Phi_{p, \theta, \delta t}[\rho] = \sum_{\mu} M^\dagger_{\mu; p, \delta t} U^\dagger_{\theta, \delta t} \rho U_{\theta, \delta t} M_{\mu; p, \delta t}
    \,.
\end{equation}
Explicitly, we have unitary rotations around $\sigma^x$ with frequency $\theta$ for total time $\delta t$, which we eventually make small
\begin{equation}
    U_{\theta, \delta t}
    =
    e^{i \frac{\theta \delta t}{2} \sigma^x}
    \,.
\end{equation}
Between the unitaries, we make weak measurements of $\sigma^z$, obtained by applying for time $\delta t$ an entangling gate with coupling strength $p$ acting between the system and an ancilla.
The weak measurement result $\mu = \pm 1$ is then obtained by a projective measurement of the ancillary degree of freedom, resulting in the Kraus operators
\begin{equation}\label{eqn:zero_dim_weak_measurement}
    M_{\mu; p, \delta t}
    =
    \frac{
    \sqrt{
    1 - p \delta t
    }
    +
    \mu
    \sqrt{p \delta t}
    \sigma^z
    }{2}
    \,.
\end{equation}
We note that we are ultimately interested in the unconditional
channel dynamics obtained by discarding the outcome $\mu$, 
but this particular unravelling represents a possible experimental realization.
Throwing away the outcomes yields a $Z$-dephasing channel $\rho \to (1-p \delta t) \rho + p \delta t \sigma^z \rho \sigma^z$.

This discrete-time dynamics is to be interpreted as a trotter approximation to the continuous-time evolution under a Lindbladian, which may be realized on a quantum simulator.  
Specifically, the continuous-time limit is defined by identifying $\rho(T)$ with the state after $T / \delta t$ total applications of $\Phi_{p, \theta, \delta t}$ while sending $\delta t \to 0$ with $\theta, p$ fixed.
\begin{equation}
    \rho(T) 
    = \lim_{\delta t \to 0}
    \Phi_{\theta, p, \delta t}^{T / \delta t}[\rho_0]
    \,.
\end{equation}
It is useful to work in the doubled Hilbert space obtained through the Choi-Jamio\l kowski isomorphism taking $\rho \to | \rho \rangle \rangle$.
Explicitly in coordinates, this sends $\rho_{ij} \ket{i}\bra{j} \to \rho_{ij} \ket{i} \otimes \ket{j}$, where in this paper, $\otimes$ denotes the tensor product between the forward and backward copies of the Hilbert space.
In this formalism, $| \rho(T) \rangle \rangle$ is obtained from the initial state $| \rho_0 \rangle \rangle$ through the time-evolution superoperator acting on the doubled Hilbert space
\begin{align}\label{eqn:cont_time_trotter}
    |\rho(T) \rangle \rangle
    &\sim
    \lim_{\delta t \to 0}
    \Big(
    e^{- \delta t 
    \frac{i \theta}{2}
    \left(
    \sigma^x \otimes I - I \otimes \sigma^x
    \right)
    }
    e^{
    - p \delta t \sigma^z \otimes \sigma^z}
    \Big)^{T / \delta t}
    |\rho_0 \rangle \rangle
    \nonumber
    \\
    &\sim \lim_{\delta t \to 0}
    \Big(
    e^{- \delta t 
    \left[
    \frac{i \theta}{2}
    \left(
    \sigma^x \otimes I - I \otimes \sigma^x
    \right)
    - p \sigma^z \otimes \sigma^z
    \right]
    }
    \Big)^{T / \delta t}
    |\rho_0 \rangle \rangle
    \nonumber
    \\
    &=
    e^{- T \mathcal{L}}
    |\rho_0 \rangle \rangle
\,,
\end{align}
where in the first line, we drop a constant prefactor which must be restored by normalizing $\Tr \rho(T) = 1$, and in the second line, we drop terms of order $\mathcal{O}\left(\delta t^2\right)$.  In the third line, the Lindbladian $\mathcal{L}$ is
\begin{equation}
    \mathcal{L} = - p \sigma^z \otimes \sigma^z + \frac{i \theta}{2} 
    \left(
    \sigma^x \otimes I - 
    I \otimes \sigma^x
    \right)
    \,.
\end{equation}
There are two salient symmetries of this model.
First, the Hermiticity-preserving property of quantum channels defines a natural $\mathcal{P}\mathcal{T}$-symmetry operator that swaps the forward and backward Hilbert spaces followed by complex conjugation.
Second, the unconditional channel dynamics has a ``weak" symmetry generated by $\sigma^x \otimes \sigma^x$.  
By ``weak" we mean that it is only a symmetry of the unconditional dynamics after discarding measurement outcomes, but not of each $M_{\mu; p, \delta t}$ independently~\cite{Buca_Prosen_Weak_Symmetry,Albert_Jiang_Weak_Symmetry}.

This motivates the definition of mutually-commuting Pauli operators
\begin{equation}\label{eqn:new_basis}
    \begin{split}
    \mu^x = \sigma^x \otimes \sigma^x &\,,\,\,
    \mu^z = I \otimes \sigma^z \\
    \tau^x = \sigma^x \otimes I &\,,\,\,
    \tau^z = \sigma^z \otimes \sigma^z \,.
    \end{split}
\end{equation}
Within this basis, the Lindbladian takes the form
\begin{equation}\label{eqn:0d_lindbladian_new_basis}
    \mathcal{L} 
    = - p \tau^z + i \frac{\theta}{2} \tau^x\left(1 - \mu^x\right)
    \,.
\end{equation}

We will be interested in the initial state $\rho_0 = \ket{0}\bra{0} = \ket{0}_\mu \ket{0}_\tau$ and consider linear probes of the time-evolved density matrix $\langle \mathcal{O} \rangle = \Tr \mathcal{O} \rho(T) / \Tr \rho(T)$.
Note that we follow the convention where $0, 1$ represents 
states in the $Z$ basis, while $+, -$ represents the $X$ basis.
The evaluation of the trace is related to the natural inner product within the doubled Hilbert space $\Tr \sigma^\dagger \rho = \langle \langle \sigma | \rho \rangle \rangle$.
Using $\sigma = I = \sqrt{2} \ket{0}_\tau \ket{+}_\mu$, we have that
\begin{equation}
\begin{split}
    \Tr \rho(T) &= 
    \sqrt{2}
    \bra{0}_\tau
    \bra{+}_\mu
    e^{- T \mathcal{L}}
    \ket{0}_\tau
    \ket{0}_\mu
    \\
    &=
    \bra{0}_\tau
    e^{- p T H_\mathrm{0d}(0)}
    \ket{0}_\tau
\end{split}
    \,,
\end{equation}
where we use $\sqrt{2} \braket{+}{0}_\mu = 1$.
Additionally, we use the fact that since $\mu^x$ is a symmetry of the dynamics, the state $\bra{+}_\mu$ fixes its value in~\eqref{eqn:0d_lindbladian_new_basis} leading to the effective Hamiltonian 
\begin{equation}\label{eqn:zero_dim_nnh}
    H_{0d}(g) = - \tau^z + i g \tau^x \,,
\end{equation}
with $g=0$.  
Similarly, the insertion of $\sigma^x = \tau^x$ within the trace leaves $\bra{+}_\mu$ unchanged, but flips the final state $\bra{0}_\tau \tau^x = \bra{1}_\tau$ such that $\Tr \sigma^x \rho(T) = \bra{1} e^{-p T H_\mathrm{0d}(0)} \ket{0}$.

The dynamics of the NHH~\eqref{eqn:zero_dim_nnh} with non-zero $g$ can be accessed by inserting $\sigma^z = \tau^z \mu^z$ or $i \sigma^y = \tau^z \mu^z \tau^x$.
In both these cases, the action of $\tau^z$ is trivial, while $\mu^z$ flips $\bra{+}_\mu \mu^z = \bra{-}_\mu$ the weak symmetry sector. 
Finally, in the case of $i \sigma^y$, this is followed by also flipping the final state.
It is most natural to consider the insertion of $\sigma^z$, since in such a case
\begin{equation}\label{eqn:0d_return_amplitude_sigmaz}
\langle \sigma^z \rangle
=
\frac{
\Tr \sigma^z \rho}
{
\Tr \rho
}= \frac{\bra{0} e^{-pT H_\mathrm{0d}(g)} \ket{0}}
{
\bra{0} e^{- p T H_\mathrm{0d}(0)} \ket{0}
}
\,,
\end{equation}
represents the return amplitude of $\ket{0}$ following imaginary-time evolution under~\eqref{eqn:zero_dim_nnh} with $g = \theta / p$.

We briefly note that in general, one may consider $\langle \mathcal{O} \rangle$, which has non-trivial time-dependence when 
$\mathcal{O} \in \mathrm{span}_{\mathbb{C}} \left\{ X, i Y\right\}$ is selected from the subspace of Pauli operators which can flip the $\bra{+}_\mu$ final state.
In particular, choosing $\mathcal{O} = a \sigma^z + i b \sigma^y$ corresponds to changing the numerator of~\eqref{eqn:0d_return_amplitude_sigmaz} to $\bra{\psi} e^{-p T H_\mathrm{0d}(g)} \ket{0}$ where $\bra{\psi} = \bra{0} a + \bra{1} b$.
Similarly, changing the initial density matrix corresponds to changing the initial state.
We note that this Pauli operator subspace is related to the ``jump operator invariant" subspace described in~\cite{Minganti_NHH_no_go}.
\subsection{Underdamped/Overdamped Transition}\label{sec:zero_dim_transition}
The behavior of $\langle \sigma^z(t) \rangle$ can be understood by diagonalizing $H_{0d}(g)$, yielding two eigenvalues $\pm \sqrt{1 - g^2}$ that switch between purely real and a purely imaginary conjugate pair at $g=1$.
The time dependence of $\langle \sigma^z(t) \rangle$ takes the generic form
\begin{equation}\label{eqn:general_zero_dim_time_behavior}
    \langle \sigma^z(t) \rangle
    =
    \frac{
    \Tr \sigma^z \rho(t)
    }
    {
    \Tr \rho(t)
    }
    \sim e^{- \Gamma t} \cos \omega t
    \,,
\end{equation}
with $\Gamma$ and $\omega$ determined from the real and imaginary parts of the eigenenergies, respectively, with distinct behavior as we vary $g$:
\begin{itemize}
    \item 
    \textbf{Overdamped Phase:}
    When $g < 1$, both eigenvalues are real, such that $\langle \sigma^z(t) \rangle$ decays exponentially and the oscillation frequency $\omega = 0$ throughout this phase.
    On the other hand, the late-time behavior is dominated by the eigenvalue with largest real part, such that the leading decay rate is given by
    \begin{equation}
        \Gamma =  p - p\sqrt{1 - g^2} \,.
    \end{equation}
    The oscillation frequency, and its derivatives $\partial_g^n \omega$ are regular as $g \to 1^-$ while the first derivative of the decay rate diverges as $\partial_g \Gamma \sim (1 - g)^{-1/2}$.
    \item
    \textbf{Underdamped Phase:}
    When $g > 1$, the eigenvalues form a purely imaginary conjugate pair and $\langle \sigma^z(t) \rangle$ exhibits damped oscillations with frequency
    \begin{equation}
        \omega = \theta \sqrt{1 - g^2}
        \,,
    \end{equation}
    vanishing as $\omega \sim (g-1)^{1/2}$ as $g \to 1^+$.
    On the other hand, $\Gamma$ is determined entirely by setting $g = 0$ for $\Tr \rho(t)$, and is constant $\Gamma = p$ throughout the phase.
    In particular, its derivatives $\partial_g^n \Gamma$ are regular.
    \item 
    \textbf{Critically Damped Point:}
    When $g =1$, our model
    lies at the so-called ``exceptional point"~\cite{Kato_Perturbation_Theory_Lin_Operators,Heiss_Physics_Exceptional_Points,Heiss_Level_Crossing,Heiss_Phases_Of_Wavefunctions,Moiseyev_NH_QM},
    the Hamiltonian~\eqref{eqn:zero_dim_nnh} fails to be diagonalizable, and has a non-trivial Jordan block structure.
    The generalized eigenmode satisfies a second-order differential equation with a non-oscillatory solution $\langle \sigma^z(t) \rangle \sim t e^{-p t}$, 
    such that 
    the decay rate formally receives a finite-time logarithmic correction
    \begin{equation}
        \Gamma = p - \frac{\log t}{t}
        \,.
    \end{equation}
    The physics at the exceptional point will not be the focus of this work.
\end{itemize}
In the non-Hermitian literature, this underdamped/overdamped transition occurs as a result of the emergence of a complex conjugate eigenvalue pair, and is associated with a spontaneous breaking of the $\mathcal{P}\mathcal{T}$ symmetry~\cite{Bender_PT_Sym,Mostafazadeh_PT_Sym,Naghiloo_YL_exp, joglekar2018passive}.
In that case, the relevant $\mathcal{P}\mathcal{T}$ operator is $\mathcal{K} \tau^z$, where $\mathcal{K}$ is complex conjugation.
Here, however, since the SWAP operator may be written $(I \otimes I + \sigma^x \otimes \sigma^x + \sigma^y \otimes \sigma^y + \sigma^z \otimes \sigma^z)/2$, we see that this $\mathcal{P}\mathcal{T}$ symmetry is nothing more than that possessed by the full Lindbladian projected into the $\mu^x=-1$ sector.
This implies that the $\mathcal{P}\mathcal{T}$-symmetry is inherited from the fundamental Hermiticity-preserving property of the quantum channel which must be preserved.  Thus, symmetry-broken states cannot be physical density matrices.
Rather, signatures of such symmetry breaking can only appear within the dynamical properties of certain observables probing the state along its relaxation to a trivial symmetric steady state.

While we have focused on the $|0 \rangle$ initial state, generic initial states will also see the underdamped/overdamped transition.  In this case, the overall scale and phase of the oscillations will be modified as a function of the size of the circle swept out on the Bloch sphere by the initial state under the unitary part of the dynamics.

Finally, we note that although we study our model in the time-continuum limit, one can also study the ``transfer matrix" acting on the double Hilbert space defined by the discrete time-evolution~\eqref{eqn:zero_dim_single_time_step}, and find identical physics and phases.
\section{Dissipation and Unitary Dynamics in a Qubit Chain}\label{sec:one_dim}
\subsection{Model Dynamics}\label{sec:one_dim_model_dynamics}
We generalize our $0$-d model to a spin chain of length $L$ 
evolving under a layer of unitaries followed by nearest-neighbor decoherence channels in each time step
\begin{equation}\label{eqn:single_layer_desc}
\begin{split}
    \Phi_{p, \theta, \delta t}&[\rho]
    \\
    =
    &\sum_{\{\mu_{ij} = \pm 1\}} 
    U^\dagger_{\theta, \delta t}
    M^\dagger_{\{ \mu_{ij} \}; p, \delta t}
    \rho
    M_{\{ \mu_{ij} \}; p, \delta t}
    U_{\theta, \delta t}
\end{split}
    \,.
\end{equation}
As in $0$-d, the unitaries consist of on-site $\sigma^x$ rotation
\begin{equation}\label{eqn:unitary_layer}
    U_{\theta, \delta t}
    =
    \prod_{i}
    e^{i \frac{\theta \delta t}{2} \sigma^x_i}
    \,,
\end{equation}
while we promote the single-site weak measurements to a 
measurement of $\sigma^z_i \sigma^z_{i+1}$ on nearest-neighbors
\begin{equation}\label{eqn:weak_measurement_layer}
    M_{\{\mu_{ij}\}; p, \delta t}
    =
    \prod_{\langle i, j\rangle}
    \frac{
    \sqrt{
    1 - p \delta t
    }
    +
    \mu_{ij}
    \sqrt{p \delta t}
    \sigma^z_i
    \sigma^z_j
    }
    {2}
    \,,
\end{equation}
where $\mu_{ij} = \pm 1$ labels the outcome.

In the continuous-time limit, we find that 
the density matrix at time $T$ can be written as 
$|\rho(T)\rangle \rangle=e^{-T \mathcal{L}} |\rho(0)\rangle \rangle$ where, up to a constant, 
the Lindbladian is
\begin{equation}\label{eqn:lindbladian_definition}
    \mathcal{L} = - p \sum_{\langle i, j \rangle}
    \sigma^z_i \sigma^z_j \otimes \sigma^z_i \sigma^z_j - i \frac{\theta}{2} \sum_i \sigma^x_i \otimes I_i - I_i \otimes \sigma^x_i \,.
\end{equation}

At each site, both the choice of unitaries and the unconditional nature of the channel gives rise to an extensive number of local weak-symmetry operators $\mu^x_j = \sigma^x_j \otimes \sigma^x_j$ which independently commute with $\mathcal{L}$.
In terms of the Pauli basis~\eqref{eqn:new_basis}, we have
\begin{equation}\label{eqn:pre_cTFIM_hamiltonian}
    \mathcal{L} = - p \sum_{\langle i, j \rangle}
    \tau_i^z \tau_j^z - i \theta \sum_i \tau_i^x \frac{1 - \mu_j^x}{2}
    \,,
\end{equation}
which makes manifest the weak $\mu^x_i$ symmetry.
As a result of these weak symmetries, $\mu_i^x$ can be viewed as a non-dynamical variable setting the magnitude of the field in~\eqref{eqn:pre_cTFIM_hamiltonian}, which 
takes the form of the complex transverse-field Ising model (cTFIM) acting on the $\tau$ spins.  
The transverse field may be turned on or off depending on $\mu_i^x$, which is set by the particular observable of interest.

\subsection{Linear Observables}\label{sec:one_dim_linear_observable}
Just as in $0$-d, the behavior of certain physical observables which are linear in the density matrix see a transition and have singular behavior as we tune $g = \theta / p$.
Following the zero-dimensional example, we consider the 
initial state $\rho_0 = (\ket{0}\bra{0})^L$.
However, we have also checked other initial states, and found the universal features of the transition, namely the diverging decay rate, to be independent of the choice of state as long as it has finite overlap with the $\mu^x_i=-1$ sector.\footnote{
We are interested in the decay rate and its characteristic divergence at long times $T \to \infty$.
In this limit, the decay rate is set by the lowest energy (in the sense of having the smallest real part) eigenstate(s) of $H_{1d}(g)$~\eqref{eqn:non_hermitian_hamiltonian} with which the initial state has a non-vanishing overlap.
For a generic initial state and generic point in parameter space $g$, this state is the ground state annihilated by all the Bogoliubov modes, as is the case for $\rho_0 = (\ket{0}\bra{0})^L$ considered here. 
}
The insertion of only a single $\sigma^z_i$ operator within the trace $\langle \sigma^z_i \rangle$ amounts to setting $\mu^x_i = -1$ and $\mu^x_i = +1$ elsewhere.
In this scenario, 
the spins at sites $j \neq i$ are fixed and serve as a bath decohering the spin at site $i$, and can effectively be traced out such that the physics is $0$-dimensional.

Therefore, the natural observable to consider is instead the expectation value of a global $\sigma^z$-string 
\begin{equation}\label{eqn:non_local_correlator}
    \langle \prod_i \sigma^z_i \rangle = 
    \frac{\Tr \prod_i \sigma^z_i \rho(T)}{\Tr \rho(T)}
    =
    \frac{
    \bra{0}^{L}
    e^{-pT H_{1d}(g)}
    \ket{0}^{L}
    }{
    \bra{0}^{L}
    e^{-pT H_{1d}(0)}
    \ket{0}^{L}
    }
    \,,
\end{equation}
where $H_{1d}(g)$ is the uniform cTFIM with field strength $g = \theta / p$
\begin{equation}\label{eqn:non_hermitian_hamiltonian}
    H_{1d}(g) = - \sum_{\langle i, j \rangle} \tau^z_i \tau^z_j
    - ig \sum_i \tau^x_i
    \,.
\end{equation}
As in Section~\ref{sec:zero_dim}, this model has a $\mathcal{P}\mathcal{T}$-symmetry~\cite{Zhang_Song_cTFIM,Sun_Tang_Kou_cTFIM,Yang_Wang_Yang_cTFIM,Tang_cTFIM,Lu_Shi_Sun_cTFIM} 
$\mathcal{K} \prod_j \tau^y_j$ descended from the Hermiticity-preserving property of the Lindbladian projected into the $\mu^x_i = -1$ local weak symmetry sector.
It was observed by the authors of~\cite{Lu_cTFIM} that there is a phase transition as $g$ is tuned associated to a ``first excited state exceptional point," but further properties of this transition and the phases it delineates were not studied.
In particular, the $\mathcal{P}\mathcal{T}$-symmetry breaking transition of
~\eqref{eqn:non_hermitian_hamiltonian} coincides with a symmetry breaking transition of the Ising $\Z_2$ symmetry $\eta = \prod_j \tau^x_j$ which 
gives rise to a long-range ordered ferromagnetic phase when the $\mathcal{P}\mathcal{T}$-symmetry is unbroken.
In contrast to the real Ising model, the other phase is gapless, has a broken $\mathcal{P}\mathcal{T}$-symmetry, and exhibits quasi-long-range order.

In the absence of decoherence, each site precesses independently about $\sigma^x$, 
and $\langle \prod_i \sigma^z_i(t) \rangle \sim \cos^L \theta T$.
On the other hand, we expect the strong-decoherence limit to have an extensive effect such that $\langle \prod_i \sigma^z_i(t) \rangle \sim e^{-p L T}$.
Away from these limits, 
we expect that $\langle \prod_i \sigma^z_i \rangle$ is still exponentially small in $L$, such that we must consider
\begin{equation}\label{eqn:l_th_root_general_form}
    \langle \prod_i \sigma^z_i \rangle^{1/L}
    \sim e^{-\Gamma T} P(T)
    \,,
\end{equation}
to obtain a reasonable thermodynamic limit.
This describes an exponentially-decaying envelope with rate $\Gamma$, modulated by a function $P(T)$ which we anticipate to be oscillatory when $g > g_c$ and non-oscillatory for $g < g_c$ for some critical $g_c$.
\begin{figure*}[t]
\centering
\includegraphics[width=1 \textwidth  ]{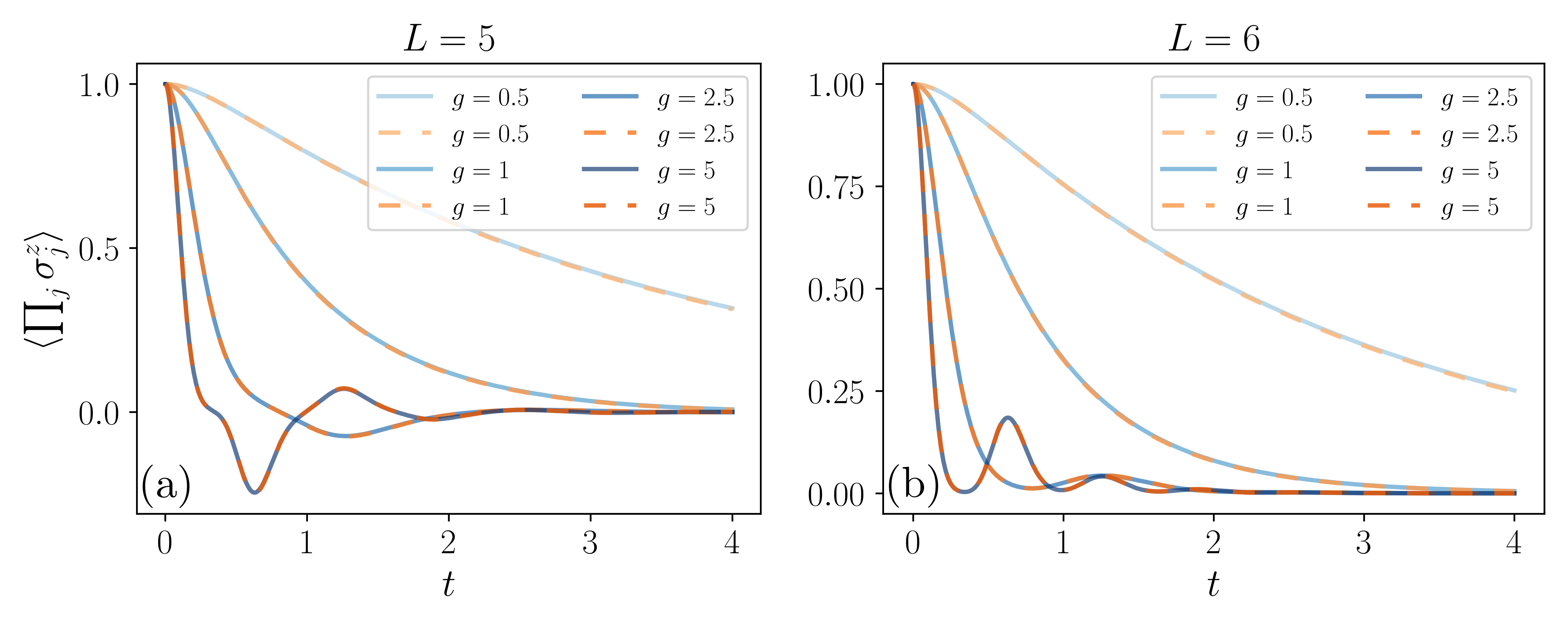}
    \caption{
    Plot of $\langle \prod_j \sigma^z_j \rangle$ for fixed $L$ and varying $g$ for the case of (a) odd $L$ and (b) even $L$.
    The solid line represents ED on the discrete-time evolution
    defined by repeated applications of
    ~\eqref{eqn:single_layer_desc} while the dashed line represents the free fermion solution~\eqref{eqn:app_observable_odd_L},~\eqref{eqn:app_observable_even_L} obtained in the continuous-time limit.
    We expect the solutions to agree when $\delta t \to 0$, 
    and already find good agreement for $\delta t = .01$ and $p = 1$.
    }
    \label{fig:ff_numerics_vary_g}
\end{figure*}

In
Section~\ref{sec:properties_one_dim_model}, 
we will derive an expression for $P(T)$, valid at late times, 
whose behavior
turns out to be rather complicated (Figure~\ref{fig:ff_numerics_vary_g}), and in fact 
depends on the parity of $L$.
On the other hand, the decay rate $\Gamma$ is well-defined regardless of the parity of $L$, 
and becomes non-analytic at the transition $g_c =1$ in the thermodynamic limit
in a way that is distinct from the zero-dimensional model.

In addition to probing the partition function, spin-spin correlation functions in the effective Hamiltonian can be probed by modifying the dynamics through the insertion of finite-strength unitaries.
We refer readers to Appendix~\ref{app:spin_correlation_probes} for details.

Finally, we note that in principle, additional terms allowed by both the local weak $\mathbb{Z}_2$ symmetry and global $\mathbb{Z}_2$ Ising symmetry are allowed in the effective non-Hermitian Hamiltonian~\eqref{eqn:non_hermitian_hamiltonian}.
Such terms may arise, for instance, from imperfect gate implementation on the quantum device.
We have also studied the model in the case that these perturbations additionally preserve the free fermion solvability of the model described in Sec.~\ref{sec:analy_one_dim}.
While non-universal features such as the oscillation frequency or the value of the critical field $g_c$ can be affected, we found that the salient universal features, namely the diverging decay rate and behavior of the spin-spin correlation functions, are unaffected by these terms.
We refer the reader to Appendix~\ref{app:perturbations} for a more detailed discussion.

\subsection{Non-Linear Observables}\label{sec:one_dim_non_linear}
The transition 
arises due to the competition between coherent unitary evolution and a decoherence channel which drives the system towards the maximally-mixed steady state at late times.
Therefore, 
it is natural to ask whether the same transition can also be probed through
basis-independent, quantum information theoretic 
observables, which are generically non-linear functions of the density matrix.

In particular, we consider the purity of the state $\Tr \rho^2(T)$ at time $T$, related to the second R\'enyi entropy $S^{(2)} = - \log \Tr \rho^2(T)$.
The purity may be computed by $\Tr \rho^2 = \Tr \rho^\dagger \rho = \langle \langle \rho | \rho \rangle \rangle$ for physical states satisfying $\rho = \rho^\dagger$.
Accounting for normalization, we have
\begin{equation}\label{eqn:purity_calc}
\frac{
    \Tr \rho^2(T) 
}
{
\left(
    \Tr \rho(T) 
    \right)^2
}
    = 
    \frac{
    \langle \langle \rho_0 |
    e^{-T \mathcal{L}^\dagger}
    e^{-T \mathcal{L}}
    | \rho_0 \rangle \rangle
    }
    {
    \left(
    \bra{0}^L
    e^{-p T H_\mathrm{1d}(0)}
    \ket{0}^L
    \right)^2
    }
    \,.
\end{equation}
Note that the numerator 
is non-trivial because the Lindbladian is not a normal operator, that is, $[\mathcal{L}, \mathcal{L}^\dagger] \neq 0$.
To evaluate it, we insert a complete set of states within the $\mu^x$ and $\tau^z$ basis.
The sum over $\mu^x$ corresponds to a sum over all possible field configurations of the effective cTFIM Hamiltonian, while the sum over $\tau^z$ inserts a resolution of the identity in the effective degrees of freedom (see Appendix~\ref{app:purity}).
One finds that
\begin{equation}
\begin{split}
    \langle \langle \rho_0 |
    &
    e^{-T \mathcal{L}^\dagger}
    e^{-T \mathcal{L}}
    | \rho_0 \rangle \rangle
    =
    \\
    &\frac{1}{2^L}
    \sum_{\{g_j = 0, g\}}
    \bra{0}^{L}
    e^{- p T H_{1d}^\dagger(g_j)}
    e^{- p T H_{1d}(g_j)}
    \ket{0}^{L}
    \,,
\end{split}
\end{equation}
which resembles~\eqref{eqn:non_local_correlator} with annealed disorder in the transverse field.
That is, $H_{1d}(g_i)$ is defined as 
\begin{equation}\label{eqn:general_1d_cTFIM}
    H_{1d}(g_i) = - \sum_{\langle i, j \rangle}
    \tau_i^z \tau_j^z - i \sum_j g_j \tau_j^x
    \,,
\end{equation}
where $g_j = g(1-\mu^x_j)/2 = 0, g$. 
We expect that the transition in the case of annealed disorder~\eqref{eqn:general_1d_cTFIM}, 
if it exists, would be determined by an effective field strength 
set by the disorder average
$g_\mathrm{eff} = \frac{g}{2}$.
Since the critical field strength in the clean case occurs at $g_\mathrm{eff} = 1$, we expect that $g_c = 2$ occurs at twice the field strength in the disordered case.
This appears to be consistent with observations in exact numerics (see App.~\ref{app:additional_observable_numerics}) of the discrete-time dynamics at small system sizes, although an analytical understanding of the existence and nature of the transition in the thermodynamic limit is left for future work.
\section{Analytic Solution in One-Dimension}\label{sec:analy_one_dim}
\setlength{\tabcolsep}{0.5em}
\begin{table*}[ht]
\begin{tabular}{|c ||c|c  |}
\hline
&&\\[-.5em]
& $L$ odd & $L$ even  \\
&&\\[-.5em]
\hline
&&\\[-.5em]
Special Momenta Modes 
& 
\makecell{
$k=0 \in \mathrm{PBC}$
\\
$k=\pi \in \mathrm{APBC}$
}
& 
\makecell{
$k=0, \pi \in \mathrm{PBC}$
\\
$k = \frac{\pi}{2} \in \mathrm{APBC}$ if $\frac{L}{4} \in \Z + \frac{1}{2}$
\\
$k = \frac{\pi}{2} \in \mathrm{PBC}$ if $\frac{L}{4} \in \Z$
\\
$\mathbf{K}, \mathbf{K'} = \mathrm{APBC}, \mathrm{PBC}$ such that $\frac{\pi}{2} \in \mathbf{K'}$
}
\\
&&\\[-.5em]
\hline
&&\\[-.5em]
Qualitative Behavior& $\langle \prod_j \sigma^z_j \rangle \sim e^{-\Gamma T} \cos \left(\omega T+ \varphi\right)$ & 
$\langle \prod_j \sigma^z_j \rangle \sim e^{- \Gamma T} + C e^{-\Gamma' T} \cos \left( \omega T + \varphi \right)$
\\
&&\\[-.5em]
\hline
&&\\[-.5em]
Oscillation Frequency 
&  
$
            \omega
            =
                \theta + 
            2p \sum_{k \in \mathrm{APBC}} \Im \epsilon_k^+
            \mathrel{\underset{L \to \infty}{=}}
            \begin{cases}
                0 &  g < 1\\
                \theta\sqrt{1 - 1 / g^2} & g > 1
            \end{cases}
            $
&
$
            \omega
            =
            \begin{cases}
                0 &  g < 1\\
                2 \theta\sqrt{1 - 1 / g^2} & g > 1
            \end{cases}
            $
\\
&&\\[-.5em]
\cline{2-3}
&&\\[-.5em]
& $\partial_g \omega \sim (g-1)^{-1/2}$ as $g \to 1^+$ when $L \to \infty$
& $\partial_g \omega \sim (g-1)^{-1/2}$ as $g \to 1^+$ for finite $L$
\\
&&\\[-.5em]
\hline
&&\\[-.5em]
Decay Rate & 
\makecell{
$\Gamma 
= \frac{p}{L} \sum_{k \in \mathrm{APBC}} 2 - \Re \epsilon_k^+ $ \\
$= \frac{p}{L} \sum_{k \in \mathrm{PBC}} 2 - \Re \epsilon_k^+$
\\
\\[-0.5em]
}& 
\makecell{
$\Gamma 
= \frac{p}{L} \sum_{k \in \bm{K}} 2 - \Re \epsilon_k^+ $ \\
$ < \Gamma' = \frac{p}{L} \sum_{k \in \bm{K'}} 2 - \Re \epsilon_k^+$ 
\\
\\[-0.5em]
}
\\
\cline{2-3}
& \multicolumn{2}{c|}{}\\[-0.5em]
& \multicolumn{2}{c|}{
\makecell{
$-\partial^2_g \Gamma \sim (g-1)^{-1/2}$ as $g \to 1^+$ when $L \to \infty$
\\
$ \partial_g \Gamma' \sim (g-1)^{-1/2}$ as $g \to 1^+$ for $L$ finite, but exponentially suppressed for $T, L$ large
\\
\\[-0.5em]
}
}
\\
\hline
\end{tabular}
\caption{Summary of our main results for $L$ odd and $L$ even.  These results are obtained in the large $T$ limit.}
\label{tab:results_summary}
\end{table*}
As shown in Sec.~\ref{sec:one_dim_linear_observable}, the expectation value of a global $\sigma^z$-string is related to the return amplitude under imaginary-time evolution of the cTFIM Hamiltonian
\begin{equation}\label{eqn:resolution_into_ghz}
\begin{split}
    \bra{0}^L
    e^{- p T H_{1d}(g)}
   &
    \ket{0}^L
    =
    \frac{1}{2}
    \bra{\mathrm{GHZ}_+}
    e^{- p T H_{1d}^{+}(g)}
    \ket{\mathrm{GHZ}_+}
    \\
    &+
    \frac{1}{2}
    \bra{\mathrm{GHZ}_-}
    e^{- p T H_{1d}^-(g)}
    \ket{\mathrm{GHZ}_-}
\end{split}
    \,,
\end{equation}
obtained by writing the initial state as 
$\ket{0}^L = \frac{1}{\sqrt{2}}(\ket{\mathrm{GHZ}_+} + \ket{\mathrm{GHZ}_-})$, and decomposing the spin Hamiltonian $H_{1d} = H^+_{1d} + H_{1d}^-$, where $H_{1d}^\pm = \Pi_\pm H_{1d} \Pi_\pm$ is the cTFIM projected to a particular Ising parity sector. 
Here, $\Pi_\pm = ( 1 \pm \eta)/2$.
We can then obtain an exact expression for the return amplitude~\eqref{eqn:resolution_into_ghz} through
the standard Jordan-Wigner transformation to a free fermion representation~\cite{Jordan_Wigner}.
We assume periodic boundary conditions for 
$H_\mathrm{1d}$, although ED with small system sizes suggest that~\eqref{eqn:resolution_into_ghz} does not depend strongly on the choice of boundary condition.
To guide our discussion, we summarize our main results in Table~\ref{tab:results_summary}.

\subsection{Qualitative Behavior}\label{sec:analytic_qualitative_behavior}
The Jordan-Wigner transform is a map between 
$H_{1d}^\pm = \Pi_\pm H_{1d} \Pi_\pm$ and a fermionic Hamiltonian, supplemented with boundary conditions determined by the parity sector.  
The fermions are anti-periodic (APBC) in the even $\eta = +1$ sector and periodic (PBC) in the odd $\eta = -1$ sector.
In Appendix~\ref{app:diagonal_free_fermion_ham}, we show that in momentum space, the resulting fermionic Hamiltonian takes the form
\begin{equation}\label{eqn:general_post_jordan_wigner}
\begin{split}
   H^\pm
   &=
   H_{k=0} + H_{k=\pi}
   \\
   &+
   2
   \sum_{\pi > k > 0}
   \Psi_k^\dagger
   \left[
   (ig- \cos k) \sigma^z
   + \sin k \sigma^y
   \right]
   \Psi_k
\end{split}
\,.
\end{equation}
Where $\Psi_k = 
\left(
\begin{smallmatrix}
    c_k \\ c_{-k}^\dagger
\end{smallmatrix}
\right)
$ is a real Majorana spinor.
The dependence on $\eta$ is manifest in the allowed momenta $k$.
When $\eta = +1$, the allowed momenta $k = \frac{\pi (2 \Z - 1)}{L}$ are odd multiples of $\frac{\pi}{L}$, 
while $k = \frac{2 \pi \Z}{L}$ are even multiples when $\eta = -1$.
As a result, $H_{k=0} = (ig - 1)( 2 c^\dagger_{k=0} c_{k=0} - 1)$ appears when $\eta = -1$, whereas the term 
$H_{k = \pi} = (ig + 1)(2 c^\dagger_{k=\pi} c_{k=\pi} - 1) $ appears with $\eta = -1$ when $L$ is even, but with $\eta = +1$ when $L$ is odd.

Away from the $k = \frac{\pi}{2}$ mode at the exceptional point $g=1$, $H^\pm$ splits into $2\times 2$ blocks labeled by momentum which can be diagonalized exactly.
The energy levels $\epsilon_k^\pm$ are
\begin{equation}\label{eqn:complex_energy_spectrum}
    \epsilon_k^\pm = \pm 2 \sqrt{1 - g^2 - 2 i g \cos k}
    \,,
\end{equation}
where the square root is evaluated using the principal branch $\Re \epsilon_k^+ > 0$.
Due to the non-Hermiticity of the Hamiltonian, associated to this complex eigenspectrum is a complete, but non-orthonormal set of eigenstates, which we use to compute the return amplitudes~\eqref{eqn:resolution_into_ghz} in Appendix~\ref{app:computing_linear_observable}.
We find that $\langle \prod_j \sigma^z_j \rangle$ can be written as a sum of two separate contributions from the APBC and PBC Brillouin zones, each of which is a product over single-particle modes.
The qualitative features of~\eqref{eqn:resolution_into_ghz} can be understood in the large-$T$ limit (see Appendix~\ref{app:long_time_limit}) which projects into the many-body ground state of~\eqref{eqn:general_post_jordan_wigner}, defined as the state with smallest real part of the complex energy.

The complex energy spectrum has a $\epsilon_k^+ \to \left(\epsilon_k^+\right)^*$ symmetry under 
the transformation $k \to \pi - k$, which effects the qualitative behavior of our observable depending on the parity of $L$.

\begin{figure*}[t]
\centering
\includegraphics[width=1 \textwidth  ]{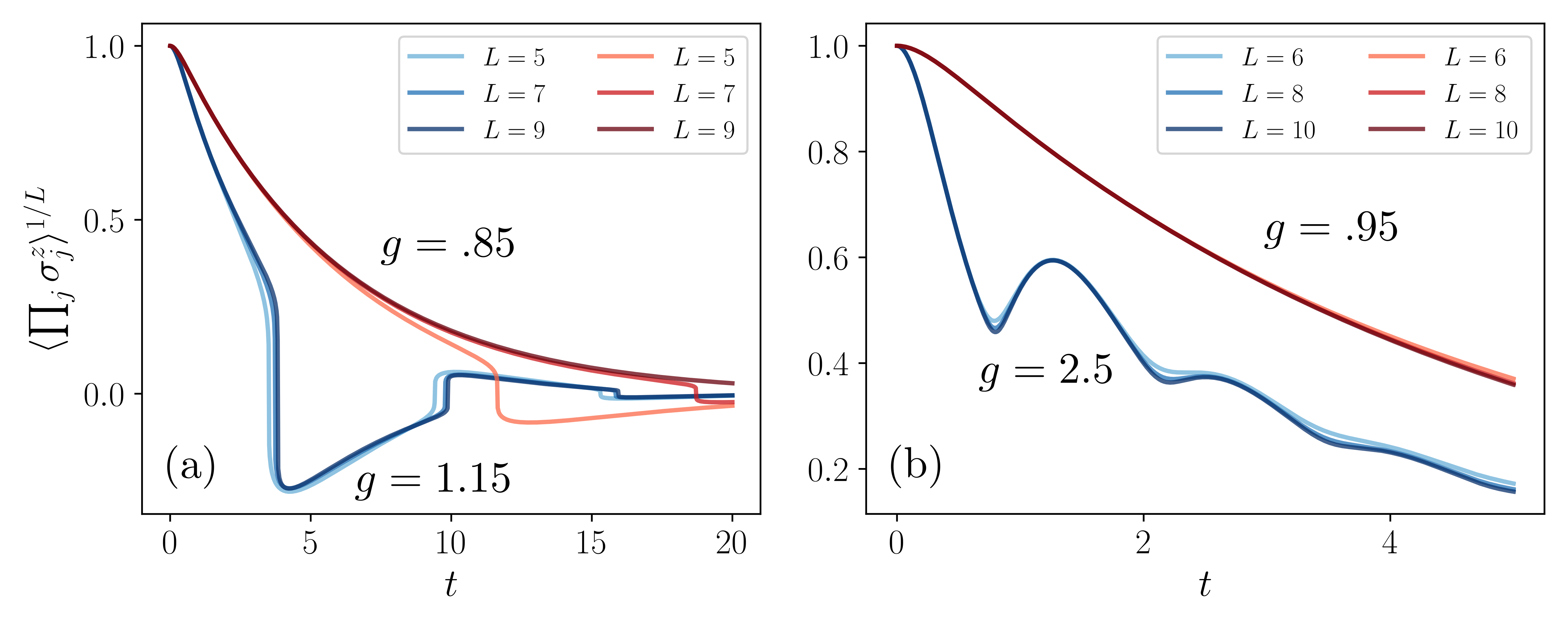}
    \caption{
    $\langle \prod_j \sigma^z_j \rangle^{1/L}$ determined by the analytic expressions~\eqref{eqn:app_observable_odd_L} and~\eqref{eqn:app_observable_even_L}, for two representative values of $g$ within the two phases for 
    varying system size with (a) $L$ odd and (b) $L$ even.  
    The $L$-th root is required to obtain a sensible thermodynamic limit~\eqref{eqn:l_th_root_general_form}.  
    The behavior is damped oscillatory~\eqref{eqn:large_t_odd_L} for $L$ odd, 
    with the frequency $\omega \to 0$ as $L \to \infty$ if $g < 1$ but 
    going to a constant for $g > 1$.  
    The behavior when $L$ is even 
    goes from a damped oscillator 
    offset by an exponentially decaying part~\eqref{eqn:large_t_even_L} for $g >1$ to purely exponential for $g < 1$, with frequency $\omega$ independent of $L$.
    }
    \label{fig:ff_numerics_vary_L}
\end{figure*}
If $L$ is odd, then 
        under $k \to \pi - k$, a momentum corresponding to APBC is taken to a PBC momentum and vice versa.
        This implies the existence of two degenerate many-body ground states with opposite parity $\eta$.
        The energies come in a complex eigenvalue pair, and in the large-$T$ limit, we show in Appendix~\ref{app:long_time_limit} that~\eqref{eqn:resolution_into_ghz} takes the form of an underdamped oscillator
\begin{equation}\label{eqn:large_t_odd_L}
    \langle \prod_j \sigma^z_j \rangle \sim e^{-\Gamma L T} \cos \left(\omega T + \varphi \right)
    \,,
\end{equation}
for all $g$, 
where the decay rate $\Gamma L$ is related to the real part of the ground state energy and the argument of the cosine is related to the imaginary part.

On the other hand, if $L$ is even, then 
        $k \to \pi - k$ takes APBC momenta to APBC momenta and PBC momenta to PBC momenta.
        In this case, the many-body ground state is unique, and is found in the parity sector that does not contain the $k = \pi /2$ mode.
        Furthermore, due to the $k \to \pi - k$ symmetry, the single-particle modes are occupied in conjugate pairs, leading to a real ground-state energy.
        The first excited state is found in the parity sector containing the $k = \pi /2$ mode, which is the symmetric point of $k \to \pi - k$.
        This implies that in the $g < 1$ underdamped phase, there is a unique first excited state, but when $g >1$ it becomes a degenerate pair with energies related by complex conjugation.
        
    In the large-$T$ limit, (Appendix~\ref{app:long_time_even_L}) we find
\begin{equation}\label{eqn:large_t_even_L}
    \langle \prod_j \sigma^z_j
    \rangle
    \sim
    e^{- L \Gamma T}
    +
    \frac{C}{2}
    e^{- L \Gamma' T}
    \cos \left( \omega T + \varphi\right)
    \,,
\end{equation}
which describes an underdamped oscillator with an exponentially decaying offset.
Here $C$ is a positive, time-independent constant and both $\Gamma$ and $\Gamma'$
are related to the real parts of the ground state and first excited state energy, respectively.

Our analytic expressions for $\langle \prod_j \sigma^z_j \rangle$ can be verified by comparison with ED numerics on small system sizes of the discrete-time dynamics~\eqref{eqn:single_layer_desc} in the limit $\delta t \to 0$.
We find good agreement in Figure~\ref{fig:ff_numerics_vary_g}.
In Figure~\ref{fig:ff_numerics_vary_L}, we verify that the time-dependence~\eqref{eqn:large_t_odd_L} and~\eqref{eqn:large_t_even_L}, strictly correct for $T \to \infty$, remain qualitatively correct at early times.  The behavior in the $L$ even case can be modified by a linear correction at the exceptional point, but we defer these details to Appendix~\ref{app:computing_linear_observable}.
The properties of the exceptional point are interesting in their own right, but contribute only a subleading effect in what follows.

\subsection{Dynamical Properties Approaching Transition}\label{sec:properties_one_dim_model}
The analytic expression for the observable $\langle \prod_j \sigma^z_j\rangle$ in the large-$T$ limit~\eqref{eqn:large_t_odd_L},~\eqref{eqn:large_t_even_L} at finite $L$ receives finite-time corrections of order $e^{-T \Delta}$, where $\Delta$ is the gap in the real spectrum.
In the overdamped phase $g<1$, the spectrum is gapped $\Delta > 2\sqrt{1-g^2}$, but closes at $k = \pi / 2$ throughout the underdamped phase $g > 1$.
In finite system sizes, the relevant gap is set by the lowest excitations $\Delta \sim \delta k \sim \frac{1}{L}$.

Therefore, our results on the nature of the transition are valid when $T \to \infty$ faster than $L \to \infty$, corresponding to 
the ordered limit $\lim_{L \to \infty} \lim_{T \to \infty}$.
The transition is characterized by the non-analytic behavior of the oscillation frequency $\omega$ and decay rate $\Gamma$, which we analyze in turn.
We find that while the oscillation frequency is determined by non-local, but effectively zero-dimensional effects, the non-analyticity of the decay rate appears only in the thermodynamic limit.

\begin{itemize}
    \item \textbf{Oscillation Frequency--}In the case of $L$ even, the single-particle modes
        come in conjugate pairs, rendering their contributions non-oscillatory.
        The $k = \pi /2$ mode, however, is unpaired and is responsible for the oscillatory part of~\eqref{eqn:large_t_even_L} with characteristic $L$-independent frequency
        \begin{equation}\label{eqn:even_freq}
            \omega_\mathrm{even}
            =
            \begin{cases}
                0 &  g < 1\\
                2 \theta\sqrt{1 - 1 / g^2} & g > 1
            \end{cases}
            \,.
        \end{equation}
        
        On the other hand, when $L$ is odd, the long-time behavior of $\langle\prod_j
        \sigma^z_j \rangle$ is always damped oscillatory, with
        frequency $\omega$ related to the imaginary part of the ground state energy
        \begin{equation}\label{eqn:discrete_sum_finite_freq}
            \omega / \theta = 1 + 
            \frac{2}{g} \sum_{k \in \mathrm{APBC}} \Im \epsilon_k^+
            \,.
        \end{equation}
        In contrast to the $L$ even case, the single-particle modes are unpaired, and their imaginary parts imperfectly cancel.
        In particular, the above sum can be evaluated in the thermodynamic limit with $\sum_k \Im \epsilon_k^+ \to \frac{1}{2} \int_0^\pi dk \, \partial_k \Im \epsilon_k^+$. 
        In the exponentially decaying phase with $g < 1$, $\Im \epsilon_k^+$ is regular and integrates to a boundary term.
        On the other hand, in the oscillatory phase $g > 1$, the integral picks
        up a delta function contribution at the branch point $k = \frac{\pi}{2}$,
        leading to a non-zero frequency
        \begin{equation}
            \omega_\mathrm{odd}
            \mathrel{\underset{L \to \infty}{=}}
            \begin{cases}
                0 &  g < 1\\
                \theta\sqrt{1 - 1 / g^2} & g > 1
            \end{cases}
            \,.
        \end{equation}

        The oscillation frequency vanishes in a non-analytic way $\omega \sim (g
        - 1)^{1 / 2}$ as $g \to 1^+$.
        The non-analyticity appears only in the thermodynamic limit for $L$ odd,
        but is present already for finite system sizes when $L$ is even.  However, it is suppressed by an exponentially small factor in $T$.
        The discrepancy between $L$ odd and $L$ even can be understood from the
        fact that the non-analyticity in $\omega$ is determined entirely by the
        exceptional point at $g= 1$ and $k = \pi / 2$ and should be thought of
        as an effectively zero-dimensional effect localized in momentum space.

        The physics of the exceptional point can be accessed through a
        particular pair of observables $\mathcal{O}_1$ and $\mathcal{O}_2$ 
        (Appendix~\ref{app:exceptional_point_probes}) with characteristic momenta $k = \pi / 2$
        such that the ratio $\Tr \mathcal{O}_1 \rho(T) / \Tr \mathcal{O}_2
        \rho(T) \sim \cos \omega_\mathrm{even} T$ exhibits infinitely long-lived
        oscillations.
        These operators exist when $L$ is even,
        since only then is $k = \pi / 2$ compatible with a periodic spin chain.
    \item \textbf{Decay Rate--}We define the decay rate $\Gamma$ at asymptotically late-times as
        \begin{equation}
            \Gamma = - \lim_{L \to \infty} \lim_{T \to \infty} \frac{1}{p L T} \log \langle \prod_j
            \sigma^z_j \rangle
            \,,
        \end{equation}
        with the caveat that for $L$ odd, we should first divide out by $\cos \omega_\mathrm{odd} T$ before the logarithm.  The late-time decay rate is related to the real part of the many-body ground state energy
        \begin{equation}\label{eqn:finite_gamma}
            \Gamma = \frac{p}{L} \sum_{\pi > k >0} 2 - \Re \epsilon_k^+
            \,.
        \end{equation}
        When $L$ is odd, the
        sum over $k$ corresponds unambiguously to either boundary conditions.
        On the other hand, when $L$ is even, we sum over $k$ corresponding to either APBC or PBC such that $k = \pi / 2$ is not included in the sum, giving us the leading ground state contribution to~\eqref{eqn:large_t_even_L}.
        $\Gamma$ and its derivatives are regular functions away from the thermodynamic limit.  However, as
        in the $0$-d model of Section~\ref{sec:zero_dim}, $\partial_g
        \Gamma'$ diverges approaching the transition from the exponential phase
        $g \to 1^-$ due to the exceptional point at $g=1$ and $k = \pi / 2$.

        In finite system sizes $\Gamma' > \Gamma$, the
        non-analyticity of the exceptional point is suppressed both
        exponentially at late times 
        and also by a factor $1 / L$
        as a measure-zero contribution to~\eqref{eqn:finite_gamma}.

        In Appendix~\ref{app:decay_rate_singularity}, we show that the leading decay
        rate $\Gamma$ diverges as $\partial_g^2 \Gamma
        \sim (g - 1)^{-1 / 2}$ as $g \to 1^+$ approaching from the oscillatory
        phase.
\end{itemize}
We remark again that the oscillation frequency $\omega$ is a non-local, but
effectively zero-dimensional effect whose non-analyticity does not have a good
thermodynamic limit, depending sensitively on UV details such as the
parity of $L$.
On the other hand, the late-time decay rate $\Gamma$ is well-defined for both
$L$ parities, and furthermore has a non-analytic behavior distinct from the $0$-d case.
For the latter, the \textit{first} derivative $\partial_g \Gamma$ diverged approaching
the transition from \textit{below} $g \to g_c^{-}$ while in the former, the
\textit{second} derivative $\partial_g^2 \Gamma$ 
diverges from \textit{above} $g \to g_c^+$.

\begin{figure*}[t]
\centering
\includegraphics[width=1 \textwidth  ]{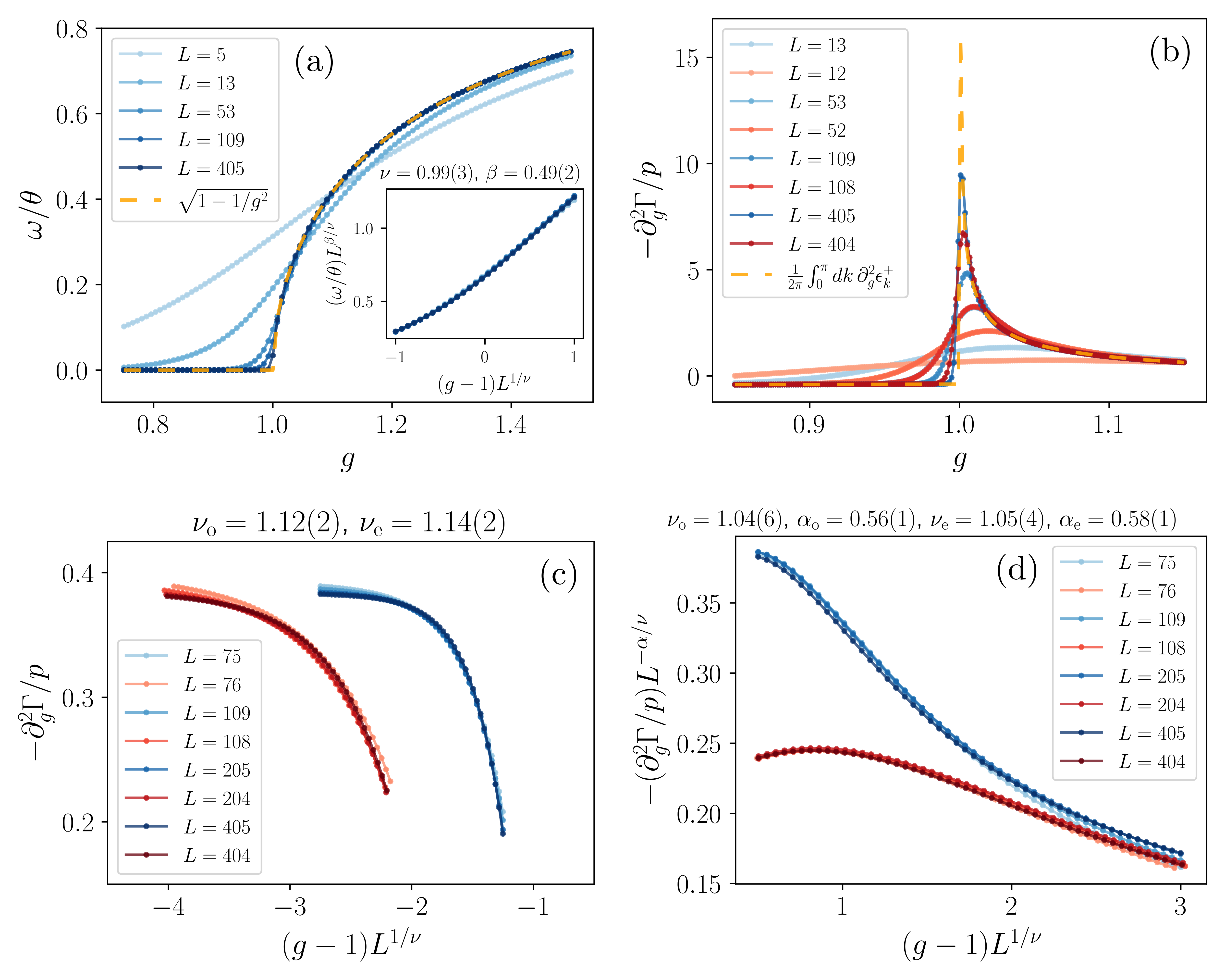}
    \caption{
    In (a), we plot the frequency $\omega / \theta$ at late times~\eqref{eqn:discrete_sum_finite_freq} for odd $L$. 
    In the inset, we collapse to $L^{-\beta / \nu}h((g-g_c)L^{1/\nu})$.
    In (b), we plot 
    $\partial_g^2 \Gamma / p$ as defined in~\eqref{eqn:finite_gamma} for $L$ both even (red colors) and odd (blue colors). 
    In (c) and (d), we do a scaling collapse of $\partial_g^2 \Gamma$ below and above the transition, respectively.
    As $g \to g_c^-$, we collapse to the scaling function $h^-_{e, o}((g-g_c) L^{1/\nu_{e, o}})$ consistent with a non-diverging order parameter at a first-order transition.
    As $g \to g_c^+$, we collapse to $L^{\alpha_{e, o} / \nu_{e, o}}h^+_{e, o}((g-g_c) L^{1/\nu_{e, o}})$ where $h^\pm_{e,o}$ are independent scaling functions.
    }
    \label{fig:scaling_collapse_late_time}
\end{figure*}

The late-time decay rate and oscillation frequency near the transition~\eqref{eqn:discrete_sum_finite_freq} and~\eqref{eqn:finite_gamma} are sums over the real and imaginary parts of the complex energy spectrum~\eqref{eqn:complex_energy_spectrum}.
The non-analytic behavior is due to the gap-closing at $k = \frac{\pi}{2}$, and expanding $\cos\left(\frac{\pi}{2} + \delta k\right) \sim \delta k \sim \frac{1}{L}$, we see that near the transition, $\epsilon_k^+$ is a scaling function of $(g-1) L^{1/\nu}$ with $\nu = 1$.
This expectation is also supported by our mean-field calculations in Section~\ref{sec:field_theory}, which suggest a first-order transition, consistent with $\frac{1}{\nu} = d = 1$~\cite{Binder_first_order,Fisher_first_order}.
These predictions are verified by a scaling collapse of the analytic late-time expressions
~\eqref{eqn:discrete_sum_finite_freq} and~\eqref{eqn:finite_gamma} in Figure~\ref{fig:scaling_collapse_late_time}.
We collapse $\omega$ to $L^{-\beta / \nu} h((g-g_c)L^{1 / \nu})$ since its expected behavior resembles a continuously vanishing order parameter near a second-order transition.
The situation for the decay rate is more delicate due to the doubled-sided nature of the transition.
Since the transition is first-order from below, we fit $\partial_g^2 \Gamma$ to
to $h^-_{e,o}((g-g_c)L^{1 / \nu_{e,o}})$.
On the other hand, $\partial_g^2 \Gamma$ is like a diverging specific heat at a second-order transition approached from above, so we fit to $L^{\alpha / \nu_{e,o}} h^+_{e, o}((g-g_c)L^{1 / \nu_{e,o}})$.
We find an even/odd ($e, o$) system size effect and generically we collapse to distinct scaling functions $h_{e,o}$ and exponents $\nu_{e,o}$.
\subsection{Numerical Study}\label{sec:one_dim_numerics}
It is a natural question to ask if our characterizations of the transition are valid away from the strict late-time limit $T \to \infty$.
To this end, we study our analytic expressions of $\langle \prod_j \sigma^z_j \rangle^{1 / L}$~\eqref{eqn:app_observable_odd_L},~\eqref{eqn:app_observable_even_L} for large, but finite $T$ and $L$, extracting $\Gamma$ and $\omega$ numerically.

\begin{figure}[t]
\centering
\includegraphics[width=.49 \textwidth  ]{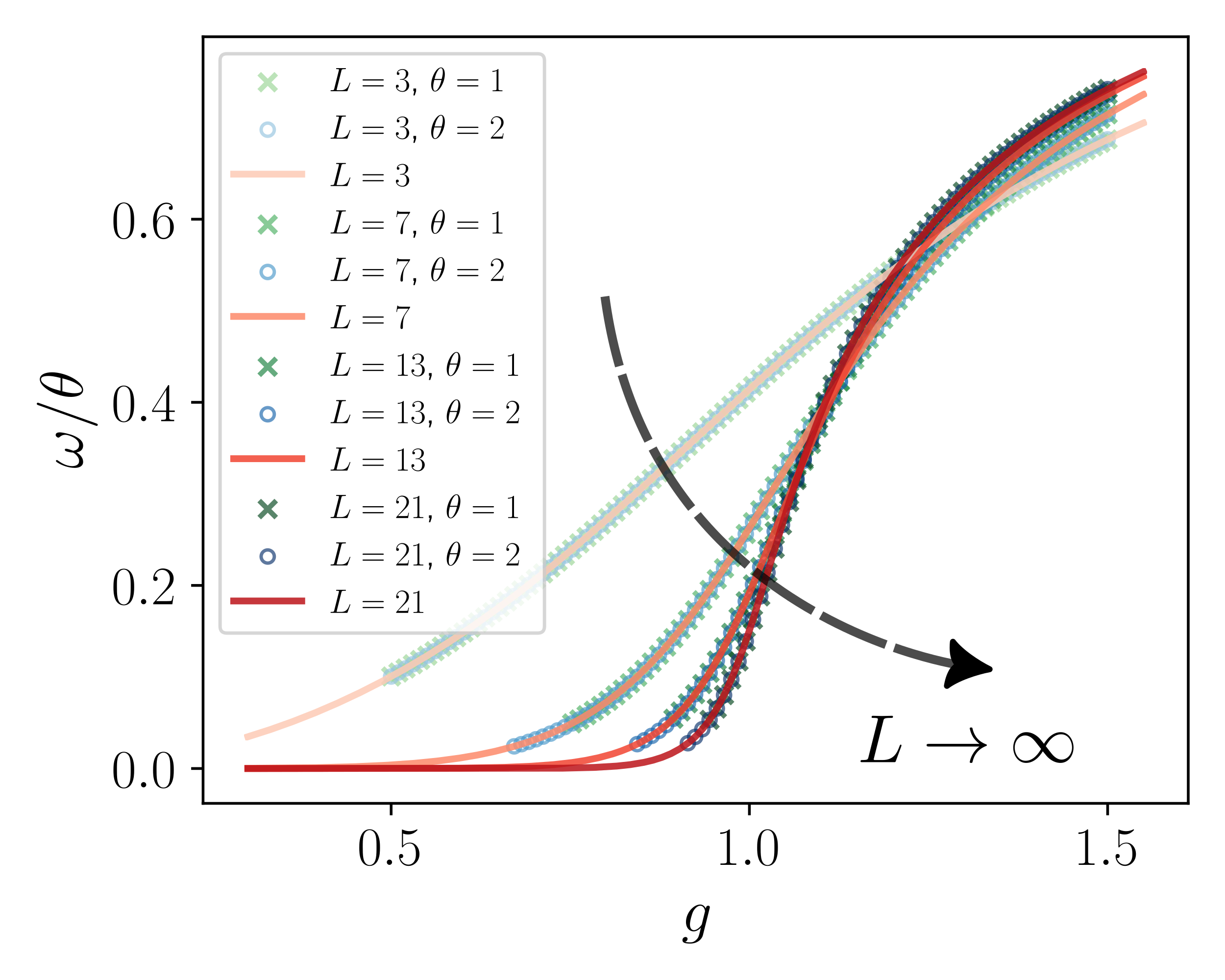}
    \caption{
    Plot of $\omega / \theta$ determined numerically from the free fermion solution for odd $L$~\eqref{eqn:app_observable_odd_L}.  We compare this to the analytic expectation in the large-$T$ limit~\eqref{eqn:discrete_sum_finite_freq}.  We find good agreement for various values of $\theta$ and $L$.
    }
    \label{fig:freq_numerics_comparison}
\end{figure}

The oscillation frequency $\omega$ can be determined most cleanly in the $L$ odd case by determining the zero-crossings of~\eqref{eqn:app_observable_odd_L}.
We find in Figure~\ref{fig:freq_numerics_comparison}, that even for not too late times $T$, the numerically determined value has good agreement with the expression~\eqref{eqn:discrete_sum_finite_freq} valid for late times.
We comment that we are limited to rather small system sizes in Figure~\ref{fig:freq_numerics_comparison}, because the numerical evaluation of $\langle \prod_j \sigma^z_j \rangle$ within the oscillatory $g > 1$ phase requires the fine phase cancellation in the product of many small factors.
This issue is most easily understood in the non-interacting unitary-only limit taking $p \to 0$ and $g \to \infty$ with $\theta = gp$ fixed.
In this limit, we have that for $L$ odd
\begin{equation}
\begin{split}
e^{i g p T}
&
    \prod_{\pi > k > 0}
    f(k) 
    =
    \\
    &
    e^{i \theta T}
    \prod_{
    m = 1
    }^{\frac{L-1}{2}}
    \cos 2 \theta T - i \sin 2 \theta T \cos \frac{\pi (2 m - 1)}{L}
\end{split}
    \,.
\end{equation}
This evaluates to the expected $\cos^L \theta T + \left( i \sin \theta T \right)^L$ using the product expansion $\frac{T_n(z) + 1}{z+1}$ for the $n$-th Chebyshev polynomial, however, numerical evaluation of the product yields an oscillatory function with twice the correct frequency.
In the case of $L$ even, the frequency follows immediately from~\eqref{eqn:app_observable_even_L} and is given by~\eqref{eqn:even_freq}.

\begin{figure*}[t]
\centering
\includegraphics[width=1 \textwidth  ]{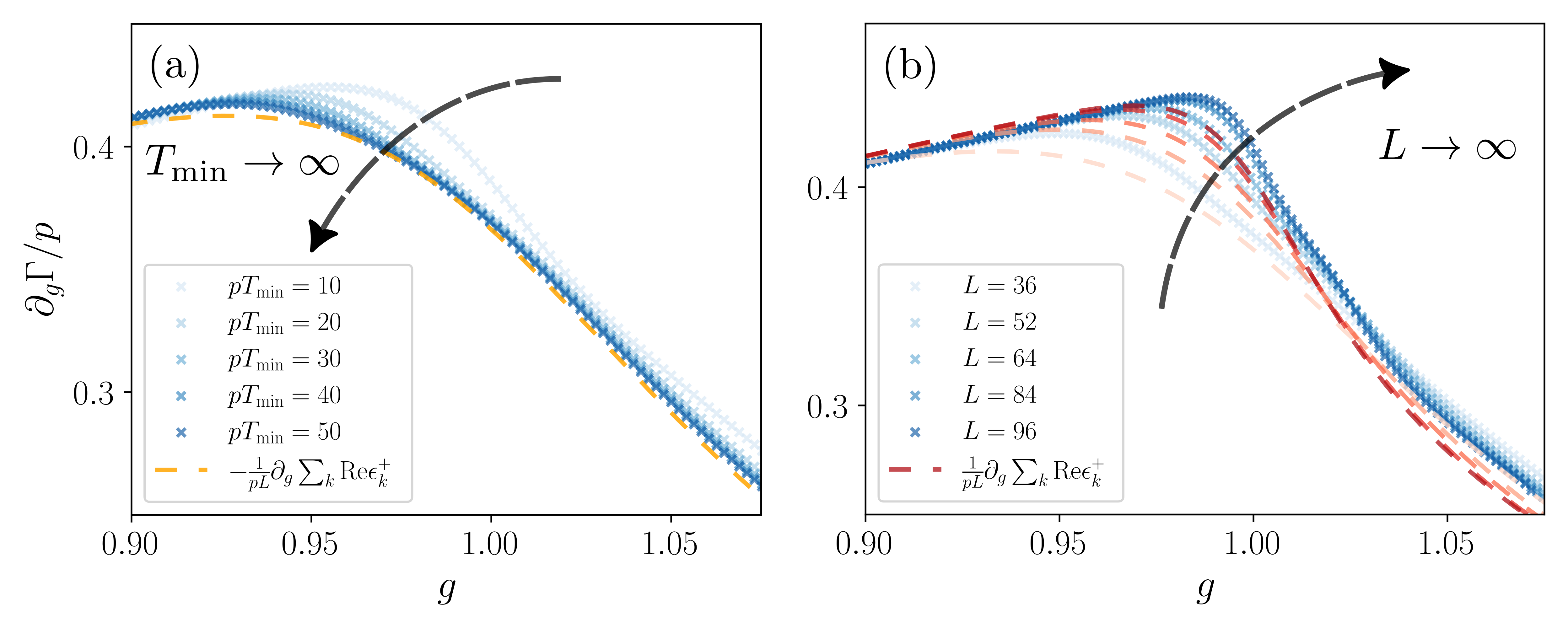}
    \caption{
    In (a), we plot $\partial_g \Gamma / p$ determined numerically from the analytic expression~\eqref{eqn:app_observable_even_L} for even $L$ by averaging over a time-window $[T_\mathrm{min}, T_\mathrm{min}+W]$ with $W = 10$, $L=32$ and $p = 1$.
    In the dashed line, we plot the expectation for late times~\eqref{eqn:finite_gamma}.
    In (b), we fix $T_\mathrm{min} = 20$ and vary $L$.
    We already find good qualitative agreement, and we expect the curves to become sharp and collapse as we take $T_\mathrm{min} \to \infty$ and $L \to \infty$.}\label{fig:decay_numerics_comparison}
\end{figure*}

Compared to $\omega$, the decay rate $\Gamma$ is numerically more stable for large $L$, but the oscillatory behavior for $L$ odd complicates its extraction within the $g > 1$ phase.
Therefore, we choose to study the $L$ even case, and determine $\Gamma$ by removing the transient oscillatory features through a simple time-averaging of $\log \langle \prod_j \sigma^z_j \rangle$ over sufficiently large interval $[T_\mathrm{min}, T_\mathrm{min} + W]$.
The derivatives $\partial^n_g \Gamma$ are then determined from $\Gamma$ through a central finite-difference approximation.
In Figure~\ref{fig:decay_numerics_comparison}, we study $\partial_g \Gamma$ and verify that it tends to the expected large-$T$ result~\eqref{eqn:finite_gamma} as $T_\mathrm{min}$ is increased with $W$ fixed.
We also study the effect of taking $L \to \infty$, and see that the peak at $g = 1$ appears to sharpen, although simultaneously taking $T$ and $L$ large is numerically difficult.

\subsection{Revealing Spin-Spin Correlations in the cTFIM in the Open-System Dynamics}\label{sec:spin_spin_correlations_1d}
Finally, we note that the dynamical phase transition within the quantity $\langle \prod_j \sigma^z_j\rangle$ can also be understood from the perspective
of the spontaneous $\Z_2$ Ising symmetry breaking of the cTFIM.
The natural probes of the transition are the spin-spin correlation functions $C(i, j) = \left(\ket{0}_H\right)^\dagger \sigma^z_i \sigma^z_j \ket{0}_H$ evaluated with respect to the ground state $\ket{0}_H$ in the right-eigenstate sense and using the regular Hermitian inner product.
In the fermionic language, such correlation functions are expectation values of non-local string operators with respect to the Bogoliubov vacuum in a particular Ising parity sector.
These can be evaluated through repeated applications of Wick's Theorem, reducing the calculation to the Pfaffian of a matrix of elementary $2$-pt correlators.

In Appendix~\ref{app:spin_correlation_functions}, we study $C(i, j)$ numerically and find that in the ``ferromagnetic" phase $g < 1$, it decays exponentially to a constant, while the ``paramagnetic" phase $g > 1$ is characterized by powerlaw correlations with a continuously-varying exponent that approaches $1/2$ at large $g$ and becomes small as $g \to 1^+$.
On the other hand, our study of the ferromagnetic phase is consistent with a non-diverging correlation length and a magnetization which jumps discontinuously to zero as $g \to 1^-$ in a way that is characteristic of a first-order transition.

Following Appendix~\ref{app:spin_correlation_probes}, the correlator $C(i, j)$ can be accessed by measuring the non-local observable $\prod_j \sigma^z_j$ at time $t=T$ with the following modified time-evolution protocol
\begin{equation}
    C(i, j) = \lim_{T \to \infty} \frac{
    \Tr
    \prod_j \sigma^z_j
    \tilde\rho(T)
    }
    {
    \Tr
    \prod_j \sigma^z_j
    \rho(T)
    }
    \,,
\end{equation}
where $\rho(T)$ is obtained by evolving the system under~\eqref{eqn:lindbladian_definition} from time $t=-T$ to $t=0$, and then evolving from $t=0$ to $t=T$ under the same dynamics but with the sign of the unitaries flipped.
Similarly, $\tilde\rho(T)$ is obtained through the same dynamics, but with the insertion of finite $e^{i \frac{\pi}{2} \sigma^z}$ unitary rotations at sites $i$ and $j$ at time $t=0$.
The infinite-time limit $T \to \infty$ is taken such that the imaginary time-evolution projects into the ground state, and we use the fact that $\left(\ket{0}_H\right)^\dagger$ is a left-ground eigenstate of $H^\dagger$.
\section{Path Integral for the cTFIM}\label{sec:field_theory}
The continuous-time formulation of our problem is amenable to field-theoretic study within the overdamped phase in higher dimensions where the Jordan-Wigner transformation cannot be applied.
The field theory can be derived by decoupling the Ising interaction with auxiliary fields $\varphi_j(\tau)$ introduced in the standard Hubbard-Stratonovich transformation and taking the continuous-time limit.
One finds that the partition function with periodic boundary conditions can be written as
\begin{equation}
    \Tr e^{- T H(g)} = 
    \int \mathcal{D}[\varphi]
    e^{-S[\varphi]}
    \,,
\end{equation}
where the $d+1$-dimensional Euclidean action is
\begin{equation}\label{eqn:euclidean_action}
\begin{split}
    S[\phi] = \frac{1}{2}
    \int d \tau
    \,
    \sum_{i , j }
    &
    \varphi_i(\tau)
    J_{ij}^{-1}
    \varphi_j(\tau)
    \\
    &- \sum_j
    \log \Tr
    \hat{T}
    e^{-\int d \tau \, h[\varphi_j]}
\end{split}
    \,,
\end{equation}
with $\hat{T}$ the time-ordering symbol and we introduce a local mean-field Hamiltonian
\begin{equation}\label{eqn:local_mean_field_hamiltonian}
    h_j[\varphi_j] = -\varphi_j(\tau) \tau^z_j - i g \tau_j^x
    \,.
\end{equation}

By varying the action~\eqref{eqn:euclidean_action}, we find the 
self-consistent mean-field condition 
$\sum_j J_{ij} \langle \tau_j^z(\tau) \rangle_{h_j[\varphi_j(\tau)]} = \varphi_i(\tau)$, where the 
expectation value is evaluated with respect to the local Hamiltonian~\eqref{eqn:local_mean_field_hamiltonian}.
Assuming a space- and time-translationally invariant solution $\varphi_j(\tau) = \phi_0$, the expectation value with respect to $h_j[\varphi_j]$ can be computed, resulting in the saddle-point condition
\begin{equation}\label{eqn:saddle_point}
    \phi_0
    \sqrt{\phi_0^2 - g^2} = 2 d 
    \phi_0
    \tanh T \sqrt{\phi_0^2 - g^2}
    \,.
\end{equation}

There are multiple solutions to~\eqref{eqn:saddle_point}.
To study the stability of these solutions, we expand the action to quadratic order in small fluctuations $\varphi(\tau) = \phi_0 + \phi(\tau)$ about the saddlepoint.
One then finds the effective action (Appendix~\ref{app:effective_action})
\begin{equation}\label{eqn:effective_action}
\begin{split}
    S_\mathrm{eff} = 
    \frac{1}{2}
    \int 
    \frac{d^d q d \omega d \nu}{(2 \pi)^{d+2}}
    \phi(q, \omega)
    \times &
    \\
    \Bigg[
    \frac{
    2 \pi \delta(\omega + \nu)}
    {
    J(\vec{q})
    }
    &-
    C(\omega, \nu)
    \Bigg]
    \phi(-q, \nu)
\end{split}
    \,.
\end{equation}
Here $J(\vec{q}) = \frac{1}{N} \sum_{i, j} e^{i \vec{q} \cdot (\vec{r}_i - \vec{r}_j)} J_{ij}$ is the nearest-neighbor Ising coupling in momentum space.
$C(\omega, \nu)$ is the Fourier transform of $C(\tau, \tau') = \langle \sigma^z(\tau) \sigma^z(\tau') \rangle_h - \langle \sigma^z(\tau) \rangle_h \langle \sigma^z(\tau') \rangle_h$,
the connected $2$-pt spin-spin correlator in time, evaluated with respect to the mean-field Hamiltonian~\eqref{eqn:local_mean_field_hamiltonian}.
\subsection{Saddlepoints}
In Appendix~\ref{app:saddlepoints_and_stability}, we find three classes of saddlepoints
\begin{itemize}
    \item \textbf{Class A--}The first class is obtained by assuming that $\phi_0 > g$.
    In the long-time limit, there is a single saddlepoint $\phi_0 = \sqrt{g^2 + 4 d^2}$ with action
    \begin{equation}
        S[\phi_0] = T N d \left[
        \left(
        \frac{g}{2d}
        \right)^2
        -1
        \right]
        \,,
    \end{equation}
    that is exponentially suppressed when $g > 2 d$ but is exponentially favored for $g < 2d$.
    By studying the effective action~\eqref{eqn:effective_action} at low energies at this saddle, we find that although the $\omega^2$ term is negative requiring us to keep higher derivatives, for all $g$, the fluctuations remain massive and with gap $\frac{1+g^2}{4d}$.
    \item \textbf{Class B--}The second class is obtained by setting $\phi_0 = 0$.
    This saddlepoint has action
    \begin{equation}
        S[\phi_0 = 0] = -  N \log 2 \cos g T
        \,,
    \end{equation}
    which is bounded from below.
    When $g > 2 d$, this saddle periodically becomes dominant as $T \to \infty$ but when $\cos g T \neq 0$.
    Fluctuations in $q$ remain massive, but the saddlepoint is unstable to time-like modulations with frequency $g / \pi$.
    \item \textbf{Class C--}The third class of saddles is obtained by setting $\phi_0 < g$ and solving the mean-field equation
    \begin{equation}
        \frac{1}{2d} \sqrt{g^2 - \phi_0^2}
        = \tan T \sqrt{g^2 - \phi_0^2}
        \,.
    \end{equation}
    When $T$ is large, the number of solutions become infinite and dense in the interval $(0, g]$.
    The action of this saddle point has the form
    \begin{equation}
        S[\phi_0 = \phi_0^*] = \frac{1}{2} \left(
        \phi_0^*
        \right)^2
        \frac{N T}{2 d}
        - N \log C(\phi_0^*)
    \,,
    \end{equation}
    where $C(\phi_0^*)$ is a $T$-independent constant.
    This saddlepoint is exponentially suppressed in the $T \to \infty$ limit, for any value of $g$.
    We also find that $q$ fluctuations are massive, but there is an instability for $\omega$ fluctuations with frequency $\sqrt{g^2 - \phi_0^2} / \pi$.
\end{itemize}
\subsection{Approaching the Transition}
Assuming that there are no other saddlepoints with smaller action, but non-trivial temporal or spatial dependence that would be missed in our analysis, we conclude that the dominant saddlepoint at late-times when $g < 2d$ is always the one of class $A$.
Furthermore,  such a saddlepoint remains stable approaching $g^* = 2d$ as fluctuations are massive for any $g$.
On the other hand, when $g > 2d$, the saddlepoints of class A and class C become exponentially suppressed for large-$T$.
The saddlepoint of class B will be dominant as long as $\cos g T$ is away from zero.
Note, however, that since such a saddle is unstable to fluctuations with frequency $g / \pi$, we might expect that the true saddlepoint is one with an oscillatory time-dependence in this phase.

These field theoretic arguments for the existence of a transition in higher dimensions are in rough qualitative agreement with our findings in the exactly solvable $1$-d model.
The fact that there is a stable, time-independent saddle for $g < g^* = 2d$ agrees with a spontaneous magnetization (Appendix~\ref{app:spin_correlation_functions}) which drops discontinuously to zero as in a first-order transition; this is also supported by the fact that both $\partial_g^2 \Gamma$ and the spin-spin correlation length do not diverge as the transition is approached from this phase.
On the other hand, the field theory is not well-understood for $g > g^*$ as it lacks a stable time-independent saddle.
It is unclear if the 
peculiar features of the $1$-d phase, namely the diverging $\partial_g^2 \Gamma$ and powerlaw spin-spin correlations, have higher-dimensional analogues.

Similarly, the fate of the $1$-d theory in the presence of generic perturbations which preserve both the local weak local and global strong $\mathbb{Z}_2$ symmetries remains unclear.
On general grounds, we expect that these perturbations will break the free fermion integrability and lead to symmetry-allowed interactions in the field theory.
In particular, it remains unclear whether the diverging decay rate $\partial^2_g \Gamma$ and powerlaw spin correlations are stable.

\section{Conclusion}\label{sec:conclusion}

In this work, we showed that there can be non-trivial phase transitions probed by linear observables of the density matrix along the approach to a trivial steady-state.
These observables are expectation values of certain non-local operators fixed by an extensive number of weak-symmetries of the open system dynamics.
By mapping to a non-Hermitian free fermion description, we derive closed analytic forms for the time-dependence of these expectation values.
The NHH has a transition between a gapped ferromagnetic phase and a gapless paramagnetic phase with powerlaw spin correlations.  We study these properties as well as the peculiar ``two-sided" nature of the transition.

One perspective on our work is that it realizes a concrete circuit dynamics exactly described by non-Hermitian physics that differs from previous constructions in that it does not rely on non-local couplings~\cite{Wei_BB_Central_Spin,Francis_Central_Spin} or post-selected trajectories~\cite{Matsumoto_YL_exp,Lee_YL_exp,Gao_YL_exp,Naghiloo_YL_exp}.
Since our approach requires locally fixing weak-symmetry sectors, it is natural to ask whether our construction can be generalized to realize any local NHH describing local degrees of freedom, or if there are any fundamental obstructions to engineering general NHHs.

On the other hand, while our construction is in principle possible on a quantum simulator with a universal gate set, it is interesting to see if similar phenomena are possible by leveraging the native interactions of a particular experimental platform where the effective low-energy degrees of freedom might not be qubit-like.
We also note that our model trades the difficulty of engineering non-local couplings in the Hamiltonian and post-selection for a non-local observable that might be experimentally realized by measuring an ancilla after it has been entangled with the rest of the system.
It is interesting to ask if there are constructions that are completely local, or if some degree of non-locality or post-selection is necessary to access non-Hermitian physics.

We note that our choice of unravelling of the decoherence channel in terms of weak measurements made by the environment~\eqref{eqn:zero_dim_weak_measurement},~\eqref{eqn:weak_measurement_layer} in the presence of local unitary rotation has a resemblance to the problem of error-correction in the $1+1$-d repetition code in the presence of coherent errors.
It was recently understood in~\cite{Fan_stab_expansion,Hauser_stab_expansion}, that the case with weak measurements and incoherent error, or projective measurements and coherent error, has a finite error-correcting threshold in the same universality class as previous results~\cite{Dennis_TC}.
However, the situation with both coherent and weak measurements remains unclear.
Numerical diagonalization at finite system sizes (see App.~\ref{app:additional_observable_numerics}) suggest similar underdamped/overdamped transitions in quantum information theoretic quantities such as the purity of the unconditional density matrix governed by a disordered cTFIM or the quantum coherent information of an ancillary qubit maximally entangled with the codespace.
These quantites computed in the unconditional dynamics lower-bound their values in the conditional dynamics relevant for error-correction; when they are close to their maximal value, it suggests that error-correction is possible.
Alternatively, the underdamped/overdamped transition may also demarcate regimes in which the coherent nature of the errors is relevant or not, resulting in decoding transitions governed by different universality classes~\cite{Dennis_TC,venn_coherent,yan_nlsm}.
The conditional dynamics represents a non-trivial generalization of our work, since the weak symmetry used in this paper is a feature only of the unconditional channel dynamics.
We leave the investigation of the conditional dynamics and its precise connection, if any, to the unconditional dynamics for future work.

Finally, our field-theoretic calculation in higher dimensions suggests a first-order transition with no diverging length scales when approached from the overdamped phase.
When the transition is approached from the underdamped phase, we find no stable, time-independent, saddlepoints. Nevertheless, we observe in the exactly-solvable $1+1$-d model a diverging second derivative of the decay rate.
Understanding if the phase transition observed in the $1+1$-d model can be understood in higher dimensions or in the presence of interactions, and in particular if the underdamped phase can be described using field-theoretic methods is an open theoretical question.

\textit{Acknowledgements:} We acknowledge helpful discussions with David M. Weld, Yaodong Li, Jacob Hauser, Andreas W.W. Ludwig, and Ana Maria Rey. S.W.Y., M.P.A.F and S.V. acknowledge support from a grant from the W. M. Keck Foundation. D. B. was supported by a Simons Investigator Award (Grant No.
511029). D.B. and M.P.A.F. were supported by the Simons collaboration on Ultra-Quantum Matter (UQM) which is funded by grants from the Simons Foundation (Grant No. 651440, 651457). D. B. acknowledges additional funding support from the National Science Foundation under Grant Number PFC PHY-2317149  (Physics Frontier Center).


\appendix

\section{Additional Observables of Non-Hermitian Hamiltonian}\label{app:nnh_mapping_operator_state}
As shown in the main text, the unconditional density matrix
$| \rho(T) \rangle \rangle$ is given by $e^{-T \mathcal{L}} | \rho(0) \rangle \rangle$, 
with Lindbladian
\begin{equation}
    \mathcal{L} = - p \sum_{\langle i, j \rangle}
    \tau_i^z \tau_j^z
    - i \theta
    \sum_j
    \tau_j^x \frac{1 - \mu_j^x}{2}
    \,.
\end{equation}
\subsection{Spin Correlation Probes}\label{app:spin_correlation_probes}
Spin correlation functions of the family of effective NHHs~\eqref{eqn:general_1d_cTFIM} can be probed through the application of single-site unitary rotations about $\sigma^z$. 
In the doubled Hilbert space, the unitary at space-time point $(r, t)$ takes the form
\begin{equation}\label{eqn:app_unitary_defect}
    \mathcal{D}(r, t) = 
    e^{- i \frac{\pi}{2} \sigma^z_r}
    \otimes
    e^{i \frac{\pi}{2} \sigma^z_r}
    = \sigma^z_r \otimes \sigma^z_r
    = \tau^z_r
    \,.
\end{equation}
Therefore, we consider the modified time-evolution determined by the insertion of $\mathcal{D}(r,t)$ at the appropriate space-time points in the continuum limit taking, $\delta t \to 0$ but keeping~\eqref{eqn:app_unitary_defect} finite.  Formally,
\begin{equation}
    \tilde \rho(T) = 
    \hat{T}
    \prod_k \mathcal{D}(r_k, t_k)
    e^{- \int_0^T d \tau \, \mathcal{L}}
    | \rho_0 \rangle \rangle
    \,,
\end{equation}
where $\hat{T}$ is the time-ordering symbol.

After inserting the appropriate non-local operator in the expectation value,
$\mathcal{L} \to H_{1d}(g)$ and $\mathcal{D}(r_k, t_k)$ inserts
the appropriate effective spin degrees of freedom
$\tau^z$ such that
\begin{equation}
\frac{
\Tr
\prod_{j=1}^L \sigma_j^z
\tilde \rho(T)
}
{
\Tr
\prod_{j=1}^L \sigma_j^z
\rho(T)
}
=
\frac{
\bra{0}^L
\hat{T}
\prod_{k}
\tau^z(r_k, t_k)
e^{- T H_{1d}(g)}
\ket{0}^L
}
{
\bra{0}^L
e^{- T H_{1d}(g)}
\ket{0}^L
}
\,.
\end{equation}
Note that the density matrix normalization is unchanged by the modified dynamics $\Tr \tilde \rho(T) = \Tr \rho(T)$.

\subsection{Exceptional Point Probes}\label{app:exceptional_point_probes}
As mentioned in the main text, when $L$ is even, it is possible to find $\mathcal{O}_1$ and $\mathcal{O}_2$ such that
\begin{equation}
    \frac{\Tr \mathcal{O}_1 \rho(T)}
    {|\Tr \mathcal{O}_2 \rho(T)|}
    \sim \cosh p \epsilon_{k=\pi / 2} T
    \,,
\end{equation}
which directly probes the non-analyticity of the exceptional point at $g=1$.
The ratio of two expectation values is required to cancel out the exponentially decaying contributions from $k \neq \frac{\pi}{2}$ that are common to both.

First, using the correspondence between operator insertions and boundary state, we know that $\bra{\tau^x = \pm 1} \mapsto \sigma^\pm = \sigma^z \pm i \sigma^y$.
Therefore, we can define the operator corresponding to the $g \to \infty$ ground state
\begin{equation}
    \mathcal{A}_\mathrm{APBC} = \prod_i \sigma^+_i \,.
\end{equation}
The corresponding operator in the PBC sector is defined by
\begin{equation}
    \mathcal{A}_\mathrm{PBC} = \frac{1}{\sqrt{L}} \sum_j \sigma^-_j \prod_{i \neq j} \sigma^+_i \,.
\end{equation}

We will also be interested in the operator corresponding to the state obtained after applying $c_{-\pi /2} c_{\pi /2}$ in the fermionic language.
The corresponding spin operator can be obtained directly from the fermionic representation 
(there will be string operators appearing, but those can be ignored after fixing an ordering of fermionic operators).
\begin{equation}
    \mathcal{B}_\mathrm{APBC} = 
    \frac{2i}{L}
    \sum_{x < y}
    (-1)^{(x-y-1)/2}
    \sigma_x^-
    \sigma_y^-
    \prod_{i \neq x, y}
    \sigma^+_i
    \,.
\end{equation}
Similarly, in the PBC sector, we define
\begin{equation}
    \mathcal{B}_\mathrm{PBC} = 
    \frac{2i}{L^{3/2}}
    \sum_{x < y}
    (-1)^{(x-y-1)/2}
    \sigma_x^-
    \sigma_y^-
    \sum_{j \neq x, y}
    \sigma_j^-
    \prod_{i \neq x, y, j}
    \sigma^+_i
    \,.
\end{equation}

We can now write 
$\mathcal{O}_1$ and $\mathcal{O}_2$ in terms of $\mathcal{A}$ and $\mathcal{B}$.
The choice between the boundary conditions (A)PBC will be determined by the parity of $L/2$; we must choose the parity sector consistent with the $k = \frac{\pi}{2}$ mode.
In order to get the purely oscillatory behavior of the $k = \frac{\pi}{2}$ mode, we should insert an equal superposition of the $n_{\pi / 2} = 0, 1$ states
\begin{equation}
    \bra{0} (P_{00} - P_{10} c_{-\pi / 2}c_{\pi / 2}) + (P_{01}^{-1} + P_{00}^{-1} c_{- \pi / 2}c_{\pi / 2})
    \,.
\end{equation}
However, we need to account for the overlap between this state and the GHZ state.
Recall that 
\begin{widetext}
\begin{equation}
\begin{split}
       \braket{\{n_{k} \}}{\mathrm{GHZ}_\pm} 
       &\supset
       \delta_{n_k, 0}
       \left(
       - i P_{00}
       \sin k/2
       -
       P_{10}
       \cos k/2
       \right)
       +
       \delta_{n_k, 1}
       \left(
       - i P_{01}^{-1}
       \sin k/2
       +
       P_{00}^{-1}
       \cos k/2
       \right)
       \\
       &\sim
       \delta_{n_k, 0}
       \left(
       - i P_{00}
       -
       P_{10}
       \right)
       +
       \delta_{n_k, 1}
       \left(
       - i P_{01}^{-1}
       +
       P_{00}^{-1}
       \right)
       \,,
\end{split}
\end{equation}
where we set $k = \frac{\pi}{2}$.
Therefore, the state we want is
\begin{equation}
    \bra{0}
    -
    \frac{P_{00} + P_{01}^{-1}}{P_{10} + i P_{00}}
    +
    \frac{P_{00}^{-1} - P_{10}}{P_{00}^{-1} - i P_{01}^{-1}}
    c_{-\pi / 2} c_{\pi / 2}
    =
    \bra{0}
     - \frac{1 - i}{\alpha_+  + i}
    +
    \frac{i \alpha_- - \alpha_+}{i \alpha_- - 1}
    c_{-\pi / 2} c_{\pi / 2}
    \,.
\end{equation}
\end{widetext}
Translating this into the operator language, we should choose
\begin{equation}
    \mathcal{O}_1 = (i - 1)(\alpha_- + i) \mathcal{A}
    + (\alpha_- + i \alpha_+)(\alpha_+ + i) \mathcal{B}
    \,.
\end{equation}
To remove the exponentially decaying components, we must divide out by the 
expectation of $\mathcal{O}_2$.
The idea is to define $\mathcal{O}_2$ such that it inserts the identical state but
with $n_{\pi / 2} = 0$ so that time evolution $e^{- p T H(g)}$ 
is a pure phase which may be removed by taking the absolute value.
Accounting for the overlap with the GHZ state, we want the state
\begin{equation}
\frac{
    \bra{0} P_{00} - P_{10} 
    c_{-\pi / 2} c_{\pi / 2}
}{
    - i P_{00} - P_{10}
}
=
\frac{
    \bra{0} 1 - \alpha_+
    c_{-\pi / 2} c_{\pi / 2}
}{
- i - \alpha_+
}
\,,
\end{equation}
so that up to a constant, we may choose
\begin{equation}
    \mathcal{O}_2
    = \mathcal{A} - \alpha_+ \mathcal{B}
    \,.
\end{equation}
Here, we have that for $g > 1$ and $k = \frac{\pi}{2}$, $\epsilon_k^+ = 2
i g \sqrt{1 - 1/g^2}$ and
\begin{equation}
    \alpha_\pm = g \left(
    1 \mp \sqrt{1 - 1/g^2}
    \right)
    \,.
\end{equation}
We note that 
this choice of operators $\mathcal{O}_1$ and $\mathcal{O}_2$ is dependent on our choice of initial state $\rho_0 = \ket{0}\bra{0}^L$ which inserts a GHZ state to the right.  
We also note that strictly speaking, the explicit form of these operators are valid in the $g > 1$ phase approaching the exceptional point from above.
A similar construction can be used at $g=1$ to determine operators $\mathcal{O}_{1, 2}$ that probe directly the linear $t$ correction~\eqref{eqn:app_generalized_schroedinger_equation} to the exponential decay due to the exceptional point.
\subsection{Purity}\label{app:purity}
We also compute the purity of the final density matrix
\begin{equation}
    \frac{
    \Tr \rho^2(T) 
}
{
    \left(
        \Tr \rho(T)
        \right)^2
}
    = \frac{
    \langle \langle \rho_0 | e^{- T \mathcal{L}^\dagger}
    e^{- T \mathcal{L}}
    |
    \rho_0 \rangle \rangle
    }
    {
    \langle \langle I | 
    e^{- T \mathcal{L}}
    |
    \rho_0 \rangle \rangle^2
    }
    \,.
\end{equation}
To evaluate the above, we insert the following resolution of the identity
\begin{equation}
    I = 
    \left( 
    \sum_{\mu^x, \tau^z = \pm 1}
        \ket{\tau^z}
        \ket{\mu^x}
        \bra{\tau^z}
        \bra{\mu^x}
    \right)^L
    \,,
\end{equation}
which has the effect of summing over all field configurations for $\mu_i^x$, in
addition to a sum over configurations for the effective $\tau$ degrees of
freedom.
Additionally, $\left(\bra{0}_\tau \bra{-}_\mu
\ket{0}_\tau \ket{0}_\mu\right)^2 = \left(\bra{0}_\tau
\bra{+}_\mu \ket{0}_\tau \ket{0}_\mu \right)^2 = 1 / 2$
tells us that each configuration is weighted by a factor $1 / 2^L$.  Therefore,
\begin{equation}
    \frac{
    \Tr \rho^2 (T) 
}{
\left( \Tr \rho(T) \right)^2}
    = 
    \frac{1}{2^L}
    \sum_{\{g_j = 0, g\}}
    \frac{\bra{0}^L 
    e^{- p T H_{1d}^\dagger(g_j)} 
    e^{- p T H_{1d}(g_j)} 
    \ket{0}^L}{
        \bra{0}^L
        e^{-2 p T H_{1d}(0)}
        \ket{0}^L
    }
    \,.
\end{equation}
\section{Underdamped/Overdamped Transition in Information-Theoretic Quantities}\label{app:additional_observable_numerics}
In this section, we numerically investigate the ``underdamped/overdamped'' phase transition in terms of quantum information theoretic probes of the density matrix $\rho(t)$ at time $t$.
\subsection{Purification Entropy}
The first quantity we chose to study is the purification entropy $\log \Tr \rho^2(t)$ starting from the initial pure state $\rho(t) = |0\rangle^{ L} \langle 0|^{ L}$.

As the state approaches the maximally mixed steady-state $\rho(t = \infty) = \mathds{1}_{2^L}$, the purification entropy will approach its maximal value $L$.  An expression for the purity was obtained in App.~\ref{app:purity}, where it is a related to a transition amplitude of a cTFIM with random field disorder.  The numerical study of the purification entropy in Fig.~\ref{fig:qinfo_probes} is consistent with the estimate of the critical $g_c \approx 2$ obtained by replacing the random field with its averaged value.
\subsection{Quantum Coherent Information}
Our model with unitary and dissipative dynamics resembles the problem of error-correction in the $1+1$-d repetition code under single-site coherent errors and repeated weak measurements.
To this end, we also study the decay of the quantum coherent information $\mathcal{I}_c$ obtained after time $t$.  The quantum coherent information is an intrinsic measure of the recoverability of information encoded within the codespace and provides a bound on the maximally achievable threshold with optimal decoding~\cite{Fan_stab_expansion}.

Concretely, for a system of size $L$, we consider the initial pure state $\rho_{SA}(0) = |\psi_0\rangle \langle \psi_0 |$ where $| \psi_0 \rangle = (|0\rangle^{L}_S |0\rangle_A + |1\rangle^{L}_S|1\rangle_A) / \sqrt{2}$ describes an ancillary qubit $A$ maximally entangled with the system $S$ in the codespace stabilized by the nearest-neighbor $\sigma^z_i \sigma^z_{i+1}$ measurements.

The $L$ qubits of the system $S$ are then evolved under the unitary and dissipative dynamics described in Sec.~\ref{sec:one_dim_model_dynamics}, while the dynamics of the ancilla $A$ are trivial.  The coherent information is then defined by
\begin{equation}
    \mathcal{I}_c(t) = S[\rho_{S}(t)] - S[\rho_{SA}(t)] \,,
\end{equation}
where $S[\rho]$ is the von-Neumann entropy of $\rho$, while $\rho_{SA}$ and $\rho_S$ are the reduced density matrices of the combined $SA$ system and $S$, respectively.

\begin{figure*}[t]
\centering
\includegraphics[width=1 \textwidth  ]{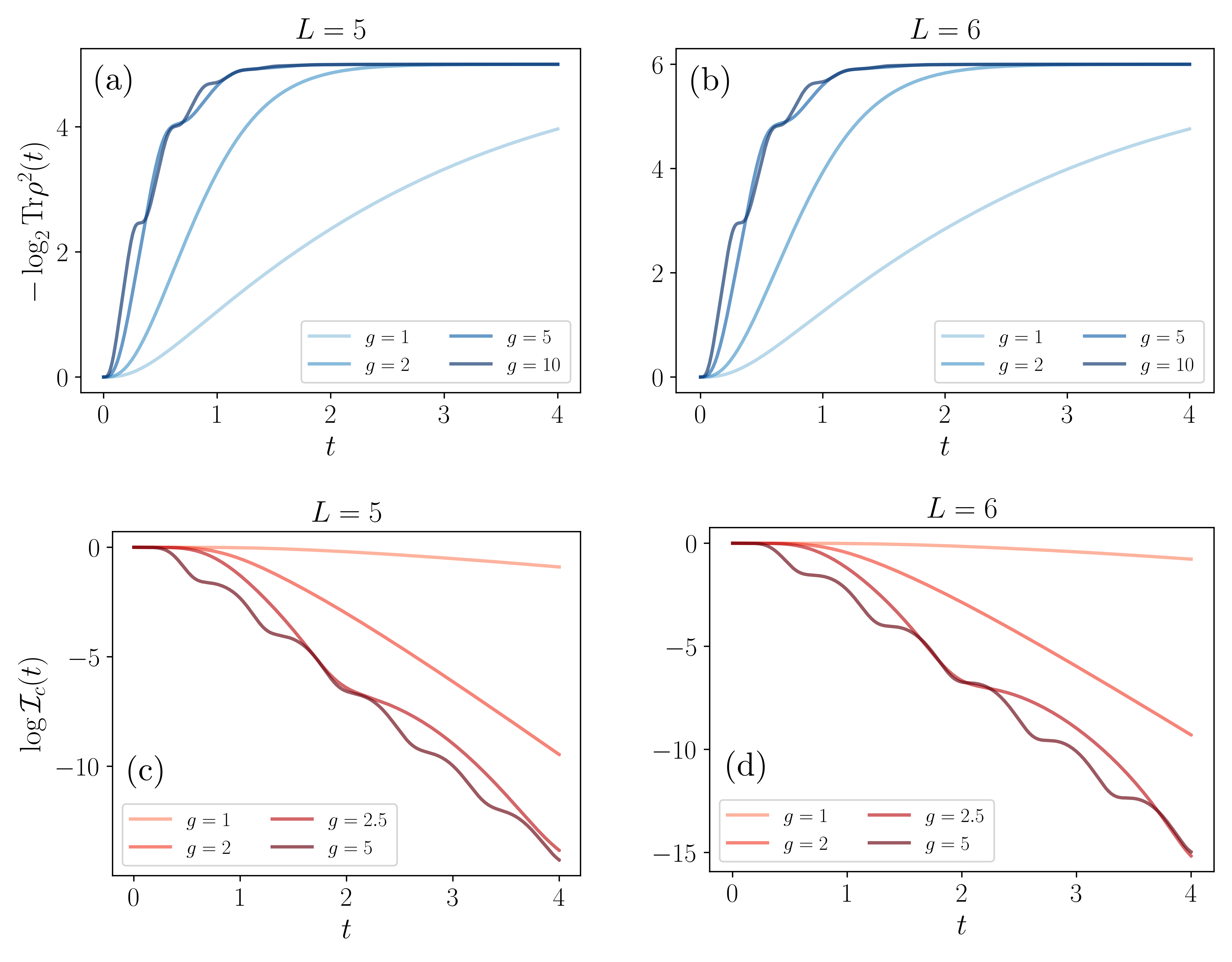}
    \caption{
    In (a) and (b), we plot the purification entropy $- \log \Tr \rho^2(t)$ for both odd and even $L$.
    In (c) and (d), we plot the quantum coherent information $\mathcal{I}_c(t)$ for both odd and even $L$.
    In all four cases, we observe an apparent phase transition between ``underdamped'' and ``overdamped'' behavior at the critical field $g_c \approx 2$.
    }
    \label{fig:qinfo_probes}
\end{figure*}

While writing down an explicit analytic form for the coherent information in terms of a transition amplitude does not seem enlightening to us at this stage, we have numerically investigated its behavior in Fig.~\ref{fig:qinfo_probes} and found that it appears to undergo an ``underdamped/overdamped'' transition at $g_c \approx 2$, coinciding with the transition in the purification entropy.
\section{Non-Hermitian Physics and Exceptional Points}\label{app:non-hermitian}
We review the formalism for diagonalizing NHHs focused within the context of our free-fermion model of the form
\begin{equation}\label{eqn:app_nnh_ff_form}
    H = \sum_{ k } \Phi_k^\dagger H_k \Phi_k \;,\;\;\;
    \Phi_k = \begin{pmatrix}
            c_k
            \\
            c_{-k}^\dagger
    \end{pmatrix}
    \,,
\end{equation}
with the $c_i$ being fermionic annihilation operators satisfying the
canonical anti-commutation relations $\{ \Phi_{k, \alpha}, \Phi_{q, \beta} \} =
0$ and $\{\Phi^\dagger_{k, \alpha} , \Phi_{q, \beta}\} = \delta_{k,
q}\delta_{\alpha, \beta}$.
Since $H \neq H^\dagger$, the Hamiltonian is generically not normal $[H, H^\dagger] \neq 0$.
Such matrices may fail to be diagonalizable, so the following discussion
depends on if the Hamiltonian sits at an exceptional point.

\subsection{No Exceptional Point}\label{app:no_exceptional_point}
Away from the exceptional point, the algebraic rank of~\eqref{eqn:app_nnh_ff_form} is equal to the geometric rank, such that $P^{-1} H P = \Lambda$ for some invertible, but not necessarily unitary matrix $P$ and diagonal $\Lambda$.
This implies that there is a set of left- and right-eigenvectors of $H$ that completely span the Hilbert space.
After diagonalization in this basis, we can write
\begin{equation}
    \Psi_k^* =
    \begin{pmatrix}
    a_k^*
    &
    a_{-k}
    \end{pmatrix}
    = \Phi_k^\dagger P
    \;,\;\;
    \Psi =
    \begin{pmatrix}
    a_{k} \\
    a_{-k}^*
    \end{pmatrix}
    = P^{-1} \Phi \,,
\end{equation}
where our notation emphasizes the fact that while $\Phi^\dagger_{k,\alpha} = \left( \Phi_{k, \alpha} \right)^\dagger$, we do not have $\Psi^*_{k, \alpha} \neq \left(\Psi_{k, \alpha} \right)^\dagger$ since $P$ is non-unitary.  
Nevertheless, invertibility is enough to preserve the anti-commutation relations $\{\Psi_{k, \alpha}, \Psi_{q, \beta}\} = 0$ and $\{\Psi^*_{k, \alpha}, \Psi_{q, \beta}\} = \delta_{k, q}\delta_{\alpha\beta}$, so $\Psi_{k, \alpha}$ and $\Psi_{k, \alpha}^*$ describe canonical creation/annihilation operators.

The time-evolution of a (left-)eigenstate $\ket{\psi_n}$ of $H$ with energy $E_n$ satisfies a first-order differential equation
\begin{equation}\label{eqn:app_schroedinger_equation}
    \left(\partial_t - E_n \right)\ket{\psi_n} = 0  \implies 
    \ket{\psi_n(t)} \sim e^{E_n t} \ket{\psi_n(t)}
    \,.
\end{equation}
\subsection{At the Exceptional Point}\label{app:at_exceptional_point}
At the exceptional point, the algebraic rank of~\eqref{eqn:app_nnh_ff_form} is larger than the geometric rank.
In the model of interest, we encounter an exceptional point of degree $2$.
This implies the existence of a unitary $U$ such that
\begin{equation}\label{eqn:app_jordan_normal_form}
    U^\dagger H U = 
    \begin{pmatrix}
        \lambda & 1 \\
        0 & \lambda 
    \end{pmatrix}
    \,,
\end{equation}
is a $2$ x $2$ Jordan block.
We can write
\begin{equation}
    \Psi_k^\dagger =
    \begin{pmatrix}
    a_k^\dagger
    &
    b_k^\dagger
    \end{pmatrix}
    = \Phi_k^\dagger U
    \;,\;\;
    \Psi =
    \begin{pmatrix}
    a_{k} \\
    b_{k}
    \end{pmatrix}
    = U^{\dagger} \Phi \,,
\end{equation}
The eigenstate $\ket{\psi^{(0)}}$ corresponding to the mode $a_k$ still satisfies the
first-order Schr\"odinger equation~\eqref{eqn:app_schroedinger_equation}.
However the generalized eigenstate $\ket{\psi^{(1)}}$ satisfies a second-order 
generalized Schr\"odinger equation
\begin{equation}\label{eqn:app_generalized_schroedinger_equation}
    \left(\partial_t - \lambda \right)^2\ket{\psi^{(1)}} = 0  \implies 
    \ket{\psi^{(1)}(t)} \sim \lambda t e^{\lambda t} \ket{\psi^{(1)}(t)}
    \,.
\end{equation}

\subsection{Generalized Bogoliubov Vacuum}\label{app:generalized_bogoliubov}
Up to normalization, the vacuum $\ket{0}_H$ with respect to $H$ is defined as the state
annihilated by all $a_k$'s away from the exceptional point, but annhilated by $a^\dagger_\mathrm{EP}$ and $b_\mathrm{EP}$ at the exceptional point.
We must occupy the $a^\dagger_\mathrm{EP}$ mode because $a_\mathrm{EP} b_\mathrm{EP} = 0$.
\begin{equation}
    \ket{0}_H = 
    a^\dagger_\mathrm{EP} b_\mathrm{EP} \prod_{k \neq \mathrm{EP}} a_k a_{-k} \ket{0}
    \,,
\end{equation}
with $\ket{0}$ the Fock space vacuum annihilated by $c_i$.  
A complete basis for the Hilbert space is generated by the states
\begin{equation}
    \ket{\{n_k\}} = 
    \left(
    a_\mathrm{EP}
    \right)^{n^a_\mathrm{EP}}
    \left(
    b^\dagger_\mathrm{EP}
    \right)^{n^b_\mathrm{EP}}
    \prod_{k \neq \mathrm{EP}}
    \left(a_k^*\right)^{n_k}
    \left(a_{-k}^*\right)^{n_{-k}}
    \ket{0}_H
    \,.
\end{equation}
Unlike the case of $H$ Hermitian, the states $\ket{\{n_k\}}$ are not
orthonormal.  The appropriate dual states are generated from the dual vacuum
\begin{equation}
    \bra{0}_H 
    = 
    \bra{0} 
    \prod_{k \neq \mathrm{EP}} 
    a^*_{-k} 
    a^*_k 
    b^\dagger_\mathrm{EP} 
    a_\mathrm{EP} 
    \,,
\end{equation}
and a complete basis for the dual Hilbert space
\begin{equation}
    \bra{\{n_k\}} = 
    \bra{0}_H
    \prod_{k \neq \mathrm{EP}}
    \left(a_{-k}\right)^{n_{-k}}
    \left(a_k\right)^{n_k}
    \left(
    b_\mathrm{EP}
    \right)^{n^b_\mathrm{EP}}
    \left(
    a^\dagger_\mathrm{EP}
    \right)^{n^a_\mathrm{EP}}
    \,.
\end{equation}
With this notation, we have the identity
$\langle
\{n_k\} | \{m_k\} \rangle = \prod_{k}\delta_{n_k, m_k}$, but we emphasize that 
$\bra{\{n_k\}} \neq \left( \ket{\{n_k\}} \right)^\dagger$.

\section{Mapping to Free Fermions}\label{app:free_fermion}
\subsection{Diagonalizing the Free Fermion Hamiltonian}\label{app:diagonal_free_fermion_ham}
The Hamiltonian $H_\mathrm{1d}(g)$ in~\eqref{eqn:non_hermitian_hamiltonian}
can be solved using the standard Jordan-Wigner mapping to free fermions.
Introducing a pair of $\gamma_i, \eta_i$ Majorana modes at each lattice site, the transformation is defined by $i \sigma^z \to \gamma \eta$, $\sigma^x \to W \gamma$ and $\sigma^y = W \eta$, 
with string operator $W_i = \prod_{j < i} i \eta_j \gamma_j$.
It is also convenient to rotate the $x/z$ axes so that $\sigma^{x/z} \to \sigma^{z/x}$ and $\sigma^y \to - \sigma^y$.
Then, in the Majorana representation, the real-space Hamiltonian takes the form
\begin{equation}\label{eqn:app_free_fermion_real_space}
    H(g) = i \sum_i \eta_i \gamma_{i+1} +  g \sum_i \gamma_i \eta_i \,.
\end{equation}
The Jordan-Wigner transform is a non-unitary transformation of the original Hilbert space,
and the fermionic representation is meant to be taken with boundary conditions
depending on the parity sector of the original spin theory.  
The Ising symmetry splits the original spin Hamiltonian into a direct sum
$H=H_{p=0} + H_{p=1}$.
Then, the parity-projected Hamiltonians
$H_{p=0, 1}$, are equivalent to the free fermion
Hamiltonian~\eqref{eqn:app_free_fermion_real_space} furnished with
anti-periodic (APBC) or periodic boundary condtions (PBC), respectively.
The case of APBC corresponds to $k \in \frac{(2 \mathbb{Z} + 1)\pi}{L}$ whereas
PBC corresponds to $k \in \frac{2 \pi \mathbb{Z}}{L}$, restricted to the first Brillouin zone $k \in (-\pi, \pi]$.
In the complex basis $c_i = \frac{1}{2} (\gamma_i + i \eta_i)$, we obtain the 
momentum space Hamiltonian
\begin{equation}
    H(g) = H_0 + H_\pi + 2 \sum_{\pi > k > 0} \Psi^\dagger_k 
    \left[
    \left(
        ig - \cos k 
        \right) \sigma^z
        + \sin k \sigma^y
    \right]
    \Psi_k
    \,,
\end{equation}
where $\Psi_k = 
\left(
\begin{smallmatrix}
    c_k \\ c_{-k}^\dagger
\end{smallmatrix}
\right)$ and the sum is over positive $k$ consistent with the boundary conditions.  
In particular, the terms
$H_0$ and $H_\pi$ may appear depending on the parity sector, and the
parity of L.
$H_0$ always appears in the odd Ising parity sector, and is absent otherwise.
$H_\pi$ appears in the odd parity sector when $L$ is even, but in the even
parity sector when $L$ is odd.  These terms have the momentum space
representation:
\begin{equation}
    H_0 = (ig - 1)(2 c^\dagger_0 c_0 - 1) \;,\;\;
    H_\pi = (ig + 1)(2 c^\dagger_\pi c_\pi - 1)    \,.
\end{equation}
As in Appendix~\ref{app:non-hermitian}, the Hamiltonian can be diagonalized away 
from the exceptional point at $k = \frac{\pi}{2}$
with
\begin{equation}
P = 
\frac{\sqrt{\sin k}}{\sqrt{\epsilon_k^+}}
\begin{pmatrix}
1 & i \\
\alpha_+ & i \alpha_-
\end{pmatrix}
\,,\;
P^{-1}
=
\frac{\sqrt{\sin k}}{\sqrt{\epsilon_k^+}}
\begin{pmatrix}
i \alpha_- & -i \\
- \alpha_+ & 1
\end{pmatrix}
\,,
\end{equation}
where 
\begin{equation}
    \alpha_\pm = \frac{2g + 2 i \cos k + i \epsilon_k^\pm}{2 \sin k} \;,
    \;\;
    \epsilon_k^\pm = \pm 2 \sqrt{1 - g^2 - 2 i g \cos k}\,,
\end{equation}
and all square roots are evaluated using the principal branch with non-negative real part.  

On the other hand, at the exceptional point, the Hamiltonian is unitarily equivalent to a Jordan normal block~\eqref{eqn:app_jordan_normal_form} with $\lambda = 0$ and
\begin{equation}
    U = \begin{pmatrix}
        1 & 1 \\
        1 & -1
    \end{pmatrix}
    \,.
\end{equation}

Away from the exceptional point, the properly normalized vacuum states satisfying $\bra{0}_H \ket{0}_H$ are
\begin{equation}\label{eqn:app_non_exceptional_point_vacuum_solution}
    \begin{split}
    \ket{0}_H &=
    \prod_{\pi > k > 0 }
    \left(
    P_{00}^{-1}
    -
    P_{01}^{-1} c_k^\dagger c_{-k}^\dagger
    \right)
    \ket{0}
        \\
    \bra{0}_H &=
    \bra{0}
    \prod_{\pi > k > 0}
    \left(
    P_{00}
    -
    P_{10}
    c_{-k} c_k
    \right)
    \end{split}
    \,,
\end{equation}
From this, we may construct a complete basis of eigenvectors of $H$ by acting upon $\ket{0}_H$ with creation operators $\ket{k} = a_k^* \ket{0}_H$.
At the exceptional point, the $k = \pi / 2$ mode is excluded from the product in~\eqref{eqn:app_non_exceptional_point_vacuum_solution} and we attach a factor $a^\dagger_{\frac{\pi}{2}} b_{\frac{\pi}{2}} \sim \frac{1}{\sqrt{2}} \left( 1 + c^\dagger_{\frac{\pi}{2}} c^\dagger_{-\frac{\pi}{2}} \right)$.
\subsection{Computing the Linear Observable}\label{app:computing_linear_observable}
We are interested in the quantity
\begin{equation}\label{eqn:app_goal_inner_product}
\begin{split}
\bra{0}^L
e^{- p T H(g)}
&
\ket{0}^L
=
\frac{1}{2}
    \bra{\mathrm{GHZ}_+}
            e^{-pT H(g)}
    \ket{\mathrm{GHZ}_+}
    \\
    &+
\frac{1}{2}
    \bra{\mathrm{GHZ}_-}
            e^{-pT H(g)}
    \ket{\mathrm{GHZ}_-}
    \,,
\end{split}
\end{equation}
where $\ket{\mathrm{GHZ}_\pm} = \frac{1}{\sqrt{2}} \left(\ket{0}^L \pm \ket{1}^L \right)$ are the Bogoliubov vacua of the $H(g=0)$ Hamiltonian in the even/odd sector
\begin{equation}\label{eqn:app_ghz_explicit_form}
    \ket{\mathrm{GHZ}_\pm}
    =
    c^\dagger_0
    \prod_{\pi > k > 0}
    \left(
    \cos \frac{k}{2} c_k^\dagger c_{-k}^\dagger - i \sin \frac{k}{2}
    \right)
    \ket{0}
    \,.
\end{equation}
The two states in~\eqref{eqn:app_ghz_explicit_form} differ in the sum over momenta, with APBC for 
$\ket{\mathrm{GHZ}_+}$, and PBC momenta for $\ket{\mathrm{GHZ}_-}$.
Also, the $c^\dagger_0$ mode (if in the odd
sector) is
always occupied while the $c_\pi$ mode is always absent.
We comment that a generic initial state will modify the details of the following analysis.
However, certain features, such as the long-time decay rate are unaffected since they depend only on the ground state~\eqref{eqn:app_non_exceptional_point_vacuum_solution}.
This has been checked in the case where $\ket{\mathrm{GHZ}_\pm}$ is replaced by the ground state of the real-transverse field Ising model away from the $g=0$ fixed point.

The GHZ states only have a non-zero overlap with states where the
$k$ and $-k$ modes are occupied/empty in pairs, i.e. $n_k = n_{-k}$.
Therefore, if we define the states $\ket{\{ n_k \}} = c^\dagger_0 \prod_{\pi > k > 0} \left(a^*_k a^*_{-k} \right)^{n_k}\ket{0}_H$ (again $c^\dagger_0$ is only present in the odd parity sector), we may compute the inner products
\begin{widetext}
\begin{equation}
    \begin{split}
       \braket{\{n_k \}}{\mathrm{GHZ}_\pm} 
       &=
       \prod_{\pi > k > 0}
       \left[
       \delta_{n_k, 0}
       \left(
       - i P_{00}
       \sin k/2
       -
       P_{10}
       \cos k/2
       \right)
       +
       \delta_{n_k, 1}
       \left(
       - i P_{01}^{-1}
       \sin k/2
       +
       P_{00}^{-1}
       \cos k/2
       \right)
       \right]
       \\
       \braket{\mathrm{GHZ}_\pm}{\{n_k \}}
       &=
       \prod_{\pi > k > 0}
       \left[
       \delta_{n_k, 0}
       \left(
       i P_{00}^{-1}
       \sin k/2
       -
       P_{01}^{-1}
       \cos k/2
       \right)
       +
       \delta_{n_k, 1}
       \left(
       i P_{10}
       \sin k/2
      k+
       P_{00}
       \cos k/2
       \right)
       \right]
       \,.
    \end{split}
\end{equation}
Substituting this into~\eqref{eqn:app_goal_inner_product}, we conclude that
\begin{align}
    \bra{\mathrm{GHZ}_\pm}
            e^{-pT H(g)}
    \ket{\mathrm{GHZ}_\pm}
    &=
    e^{- p T C}
    \sum_{
    \{ n_k \}
    ,
    \{ m_k \}
    }
       \braket
       {\mathrm{GHZ}_\pm} 
       {\{n_k \}}
       \bra{\{n_k \}}
       e^{-p T H(g)}
       \ket{\{m_k \}}
       \braket
       {\{m_k \}}
       {\mathrm{GHZ}_\pm} 
    \\
    &=
    e^{- p T C}
    \sum_{
    \{ n_k \}
    }
       \braket
       {\mathrm{GHZ}_\pm} 
       {\{n_k \}}
       \braket
       {\{n_k \}}
       {\mathrm{GHZ}_\pm} 
       e^{-pT \sum_k (2 n_k - 1) \epsilon_k^+}
       \\
       &=
    e^{- p T C}
       \prod_{\pi > k > 0}
       \left[
       \cosh \left(
       p T \epsilon_k^+
       \right)
       +
       \frac{2}{\epsilon_k^+}
       \left(
       1 - i g \cos k
       \right)
       \sinh \left(
       p T \epsilon_k^+
       \right)
       \right]
       \,,
\end{align}
\end{widetext}
where in the first line, we inserted a resolution of the identity within the eigenbasis of $H(g)$, in the second line we use the fact that $H(g)$ is diagonal in that basis, and finally in the last line we simplify the expression using the explicit forms of the inner product between $\ket{\mathrm{GHZ}_\pm}$ and the basis states $\ket{\{n_k\}}$.
Here $C$ includes the contribution from $H_{0/\pi}$, and its value depends on the parity sector and parity of $L$.

In particular, note that $\bra{\mathrm{GHZ}_\pm}e^{-pTH(0)}\ket{\mathrm{GHZ}_\pm} = e^{-2 p T L}$.  This implies that
\begin{widetext}
\begin{equation}\label{eqn:app_linear_quantity_general_L}
    \langle
    \prod_i
    \sigma^z_i
    \rangle
    =
    \frac{
    \bra{0}^L
    e^{-pT H(g)}
    \ket{0}^L
    }{
    \bra{0}^L
    e^{-pT H(g=0)}
    \ket{0}^L
    }
    =
    \frac{1}{2}
    \left(
    e^{-p T C_{\mathrm{APBC}}}
    \prod_{k \in \mathrm{APBC}}
    f(k)
    +
    e^{-p T C_{\mathrm{PBC}}}
    \prod_{k \in \mathrm{PBC}}
    f(k)
    \right)
    \,,
\end{equation}
\end{widetext}
where $f(k)$ is defined by
\begin{equation}\label{eqn:app_f_function}
   f(k) = \frac{
       \cosh \left(
       p T \epsilon_k^+
       \right)
       +
       \frac{2}{\epsilon_k^+}
       \left(
       1 - i g \cos k
       \right)
       \sinh \left(
       p T \epsilon_k^+
       \right)
   } 
   {
        e^{2pT}
   }
   \,.
\end{equation}
The behavior of~\eqref{eqn:app_linear_quantity_general_L} depends on the parity of $L$.
At the exceptional point, $f\left(\pi / 2\right)$ is replaced by
\begin{equation}
    f_{g = 1}\left(
    \frac{\pi}{2}
    \right)
    =
    \frac{1}{2} \left(
    1 - p T
    \right)
    e^{- 2 p T}
    \,.
\end{equation}
\subsection{Dependence on the Parity of L}\label{app:L_parity}
Recall that momenta in the Brillouin zone take the form $\frac{n \pi }{L}$, where $n \in 2 \mathbb{Z} + 1$ is odd for APBC and even $n \in 2 \mathbb{Z}$ for PBC.
Under the transformation $k \to \pi - k = \frac{\pi (L - n)}{L}$, when $L$ is odd, $L - n$ is of opposite parity so a momentum corresponding to APBC is taken to a PBC momentum and vice versa.
Additionally, note that $f(k)$ has the property $f(\pi - k) = f^*(k)$.  Thus, we conclude that the two products in~\eqref{eqn:app_linear_quantity_general_L} are complex conjugates 
\begin{equation}\label{eqn:app_observable_odd_L}
    \langle
    \prod_j \sigma^z_j 
    \rangle
    =
    \Re
    \left(
    e^{i g p T}
    \prod_{\pi > k > 0}
    f(k)
    \right)
    \,,
\end{equation}
where $k = \frac{\pi (2n-1)}{L}$ with $n = 1, 2, \dots, \frac{L-1}{2}$.

When $L$ is even, the transformation $k \to \pi - k$ now takes momenta in APBC to a different momenta in APBC.
Similarly, it takes a momenta in PBC to another momenta in PBC.
Because each $f(k)$ comes with a corresponding factor of $f^*(k)$ in the product, we may replace $\prod_k f(k) \to \prod_k |f(k)|$.
This will be true for all momenta modes other than the special symmetric mode
$k=\frac{\pi}{2}$, which is invariant under $k \to \pi - k$.  
When $g \neq 1$, this mode contributes
\begin{equation}\label{eqn:app_symmetric_mode_form}
\begin{split}
    e^{2p T} f
    \left(
        \frac{\pi}{2}
    \right)
    &=
    \cosh
    \left(
    2p T 
    \sqrt{1 - g^2} 
\right)
\\
&+
\frac{1}{\sqrt{1 - g^2} }
\sinh
    \left(
    2p T 
    \sqrt{1 - g^2} 
    \right)
    \,,
\end{split}
\end{equation}
and can be negative when $g > 1$.  The $k = \frac{\pi}{2}$ mode can belong to either the APBC or PBC sector, depending on the parity of $L/2$.
We will assume that $L/2$ is even such that $k = \frac{\pi}{2}$ belongs to the PBC sector, but the other case is similar.
In that case we have: 
\begin{equation}\label{eqn:app_observable_even_L}
\langle
\prod_j \sigma^z_j 
\rangle
=
\frac{1}{2}
\prod_{\pi > k > 0}
|f(k)|
+
\frac{1}{2}
\sgn 
\left[
f \left(
    \frac{\pi}{2}
\right)
\right]
\times
\prod_{\pi > k' > 0}
|f(k')|
\,,
\end{equation}
with $k = \frac{\pi (2n-1)}{L}$ with $n = 1, 2, \dots, \frac{L}{2}$ corresponding to APBC and $k' = \frac{2 \pi n}{L}$ with $n=1, 2, \cdots \frac{L}{2} - 1$ corresponding to PBC and including $k=\frac{\pi}{2}$.
When $L/2$ is odd, the labels of $k$ and $k'$ are flipped.

\section{Long-Time Behavior}\label{app:long_time_limit}
In the limit $T \to \infty$, the form of $f(k)$~\eqref{eqn:app_f_function} simplifies greatly.  Assuming that $k \neq \frac{\pi}{2}$, such that $\Re \epsilon_k^+ > 0$ is strictly positive, we find that
\begin{equation}\label{eqn:app_large_T_f_k_form}
    f(k) 
    \to
    \frac{1}{2}
    e^{pT(R- 2)}
    e^{i p T I(k)}
    \left[
        1 + \frac{2(1 - i g \cos k)}{R + i I}
    \right]
    \,,
\end{equation}
where $R(k)$ and $I(k)$ represent the real and imaginary parts of $\epsilon_k^+$, respectively.  
\subsection{Case of Even L}\label{app:long_time_even_L}
Recall that for even $L$, the observable $\langle \prod_i \sigma^z_i
\rangle$ is given by the average of two products~\eqref{eqn:app_observable_even_L}.
The form of $f(k)$ at large $T$~\eqref{eqn:app_large_T_f_k_form} tells us that
\begin{equation}
    \prod_{\pi > k > 0} |f(k)| \to 
    e^{- pT \sum_k (2 - R(k))}
    \prod_{\pi > k > 0}
    \Big|
        1 + \frac{2(1 - i g \cos k)}{R + i I}
    \Big|
    \,,
\end{equation}
and 
\begin{equation}
\begin{split}
    \prod_{\pi > k' > 0} |f(k')| \to
    &
    e^{- pT \sum_{k'} (2 - R(k') )}
    \times
    e^{2 p T}
    \Big|
f \left(
    \frac{\pi}{2}
\right)
    \Big|
    \\
    &\times
    \prod_{
    \substack{
    \pi > k' > 0
    \\
    k' \neq \pi / 2
    }
    }
    \Big|
        1 + \frac{2(1 - i g \cos k')}{R + i I}
    \Big|
    \,.
\end{split}
\end{equation}
The large-$T$ behavior will be determined by the argument of the exponential prefactor.
In particular, one finds that
\begin{equation}
    \sum_{\pi > k > 0} 2 - R(k)  < 
    \sum_{\pi > k' > 0} 2 - R(k') 
    \,,
\end{equation}
that is, the product containing $k' = \frac{\pi}{2}$ will be subleading at late-times.
This is true even at the exceptional point, since the linear correction to $f(\pi / 2)$ is exponentially surpressed by the product over $k'$ at sufficiently late times.
The leading contribution to $\langle \prod_i \sigma^z_i \rangle$ then comes from the product over $k$, which decays exponentially with rate
\begin{equation}\label{eqn:app_decay_rate_exp}
    L \Gamma = 
    \sum_{\pi > k > 0} 2 - R(k)
    \sim
    \frac{p L}{2 \pi}
    \int_0^\pi
    2 - R(k)
    \,,
\end{equation}
where we've taken the $L \to \infty$ limit to convert the sum into an integral.  

On the other hand, the product over $k'$ containing the $\frac{\pi}{2}$ mode is subleading, and has the form of an underdamped oscillator $e^{-\Gamma'  L T} \cos \omega T$ when $g > 1$.
For finite $L$, $\Gamma' > \Gamma$, but $\Gamma' \to \Gamma$ in the thermodynamic limit.
The oscillation frequency
$\omega = 2 \theta \sqrt{1 - 1 / g^2}$ 
is determined entirely by~\eqref{eqn:app_symmetric_mode_form},
which has a square root singularity at the transition point even for finite $L$.

\subsection{Case of Odd L}\label{app:long_time_odd_L}
In the case of odd $L$, the observable $\langle \prod_i \sigma^z_i \rangle$ behaves as an underdamped oscillator and can be written as
\begin{equation}
    \langle
    \prod_i \sigma^z_i \rho(T) 
    \rangle
    \sim
    e^{- p T \sum_k 2 - R(k)}
    \cos \Theta \,,
\end{equation}
where we've substituted the large-$T$ limit of $f(k)$~\eqref{eqn:app_large_T_f_k_form} into~\eqref{eqn:app_observable_odd_L}.  Here $\Theta$ is given by
\begin{equation}
\begin{split}
    \Theta  
    &=
    \arg
    \left(
    e^{i \theta T}
    \prod_{k \in \mathrm{APBC}} f(k)
    \right)
    \\
    &= T
    \left(
    \theta + p \sum_{k \in \mathrm{APBC}} I(k)
    \right) + \varphi(g, L)
    \,,
\end{split}
\end{equation}
where $\varphi(g, L)$ is a time-independent phase given by
\begin{equation}
    \varphi = \sum_{ \pi > k > 0} \arg \left(
    1 + \frac{2- 2 i  g \cos k}{R + i  I}
    \right)
    \,.
\end{equation}
The decay rate $\Gamma$ is given by the same expression as for $L$ even~\eqref{eqn:app_decay_rate_exp}.  On the other hand, the phase $\Theta$ determines the oscillation frequency $\omega = \theta + p \sum_k I(k)$.
\begin{align}\label{eqn:app_odd_L_sum_def_frequency}
        \omega  / \theta 
               = 
                   1 &+ 2 \sum_{k \in \mathrm{APBC}}
                   \Im
                   \left(
                   \sqrt{\left(\frac{1}{g}\right)^2 - 1 -\frac{2i}{g}\cos k} 
                   \right)
                   \nonumber
               \\
               =
                   1 &+ \sum_{k \in \mathrm{APBC}}
                   \Im
                   \left(
                   \sqrt{\left(\frac{1}{g}\right)^2 - 1 -\frac{2i}{g}\cos k} 
               \right)
               \nonumber
               \\
               \;\;\;\;\;
                   &- 
                   \sum_{k \in \mathrm{PBC}}
                   \Im
                   \left(
                   \sqrt{\left(\frac{1}{g}\right)^2 - 1 -\frac{2i}{g}\cos k} 
                   \right)
    \,,
\end{align}
where we used $I(\pi - k) =
-I(k)$.
The two sums above can be combined into a single sum of terms of the form
$h(k + \pi / L) - h(k) \approx \frac{\pi}{L} \partial_k h(k)$.  Additionally, in the large-$L$ limit, we have $\sum_k \to \frac{L}{2\pi} \int_0^\pi dk$ such that
\begin{align}
    \omega / \theta
    \approx
        1 &- 
        \frac{1}{2}
        \int_0^\pi
        dk \,
        \partial_k 
                   \Im
                   \left(
                   \sqrt{\left(\frac{1}{g}\right)^2 - 1 -\frac{2i}{g}\cos k} 
               \right)
               \nonumber
    \\
    =
        1 
        &- 
        \frac{1}{2}
        \Im 
        \left(
            \sqrt{\left(\frac{1}{g}\right)^2 - 1 + \frac{2i}{g}} 
        \right)
        \\
        \;\;\;\;\;
        &+ 
        \frac{1}{2}
        \Im 
        \left(
            \sqrt{\left(\frac{1}{g}\right)^2 - 1 - \frac{2i}{g}} 
        \right)
        - \frac{1}{2} \Delta_{\pi / 2}
               \nonumber
    \,,
\end{align}
where $\Delta_{\pi / 2} = \lim_{\delta \to 0} \Im \sqrt{1 / g^2 - 1 - 2 i \cos (k + \delta) / g} - \Im \sqrt{1 / g^2 - 1 - 2 i \cos (k - \delta) / g}$ accounts for the possible delta function discontinuity at $k = \frac{\pi}{2}$ that occurs due to the branch cut for $g >1$.  When $g < 1$, $\Delta_{\pi / 2} = 0$.  Again picking the principal branch cut with positive real part, we find that $\Im \sqrt{1 / g^2 - 1 + 2i / g}  = -\Im \sqrt{1 / g^2 - 1 - 2i / g} = -1$, and $\Delta_{\pi / 2} = 2 \sqrt{1 - 1 / g^2}$.  Therefore, we conclude that the in the large $T, L$ limit, the oscillation frequency behaves as
\begin{equation}
    \omega  = \theta \sqrt{1 - \frac{1}{g^2}}
    \,.
\end{equation}
Note that this is half the frequency of the $L$ even case.  Additionally, note that when $L$ is even, the non-analytic square root behavior appears for any finite $L$, but the thermodynamic limit is required for $L$ odd. 
This is because the singular behavior is determined entirely by the $k = \frac{\pi}{2}$ mode, which is only compatible with $L$ even.
For finite $L$ odd, the frequency is instead given by the sum~\eqref{eqn:app_odd_L_sum_def_frequency}, which is analytic away from the thermodynamic limit.

\subsection{Singularity in the Decay Rate}\label{app:decay_rate_singularity}
The decay rate for $L$ both odd and even is given by the finite sum
\begin{equation}
    \Gamma
    = \frac{p}{L} \sum_{\pi > k >0} 2 - R(k)  
    \,,
\end{equation}
where the sum over $k$ avoids the symmetric point $k = \frac{\pi}{2}$ and therefore is an analytic function at the transition point $g_c = 1$ for any finite $L$.  
We claim that in the thermodynamic limit, the second derivative $\partial_g^2 \Gamma$ diverges at the critical point.
Indeed, for large $L$, we may convert the above sum into an integral  
\begin{align}\label{eqn:gamma_diverge_sum_def}
    \partial_g^2 \frac{\Gamma}{p}
    &=
    -
    \frac{1}{L}
    \sum_k
    \partial_g^2
    R(k)
    \nonumber
    \\
    &\approx 
    - \frac{1}{2\pi}
    \int_0^\pi
    dk \,
    \partial_g^2
    R(k)
    =   
    - \frac{1}{2\pi}
    \int_0^\pi
    dk \,
    \partial_g^2
    \epsilon_k
    \nonumber
    \\
         &=
    - \frac{1}{2 \pi}
    \int_0^\pi
    dk \,
     \frac{2 \sin^2 k}{\left( 1 - g^2 - 2 i g \cos k\right)^{3/2}}
    \nonumber
     \\
     &\approx
     -
    \frac{1}{2 \pi}
    \int_{-\Lambda}^\Lambda
    \frac{dk}{\left(1 - g^2 -2 i g k\right)^{3/2}}
    \,,
\end{align}
where in the first line, we used the property $\epsilon_k^+ = \left( \epsilon_{\pi - k}^+\right)^*$ and in the third line we expand the integral around the singularity at the branch point $k = \frac{\pi}{2}$.

There are two cases:
\begin{itemize}
    \item $g<1$:  In this case, the argument of $\left(1 - g^2 + 2 i g k\right)^{3/2}$ is $\frac{3}{2}\arctan\left(\frac{2 g k}{1-g^2}\right)$, which is a smooth function of $k$ near $k=0$.  Therefore, we may directly evaluate the integral, after which we find
    \begin{align}
    \partial_g^2
    \frac{\Gamma}{p}
    &\sim
        - \frac{i}{2 \pi g} \frac{
        \sqrt{1-g^2 + 2 i g \Lambda}
        -
        \sqrt{1-g^2 - 2 i g \Lambda}
        }{\sqrt{\left(1 - g^2\right)^2 + 4 g^2 \Lambda^2}}
        \nonumber
        \\
        &\underset{g \to 1^-}{=}
        \frac{1}{2 \pi \sqrt{\Lambda}}
        \,,
    \end{align}
    which is regular as $g \to 1^-$.
    \item $g > 1$:  In this case, because of the branch cut along the negative real axis, the argument of $\left(1 - g^2 + 2 i g k \right)^{3/2}$ takes the form
    $\frac{3}{2} \arctan \left( \frac{2 g k}{1-g^2}\right) - \frac{\pi}{2} \sgn k + 2 \pi \sgn (k) \Theta \left( \Big| \frac{2 g k }{1 - g^2} \Big|- \sqrt{3}\right)$, with $\Theta$ the Heaviside step function.
    In this case, the integral needs to be done in parts, $k > 0$ and $k < 0$ to avoid the branch cut.  If we define the anti-derivative
    \begin{equation}
        F(k) = \frac{i}{2 \pi g} \frac{1}{\sqrt{1-g^2 + 2 i g k}} \,,
    \end{equation}
    then the integral has the value
    \begin{align}
    \partial_g^2
    \frac{\Gamma}{p}
        &\sim - 2 i \left( F(\Lambda) - F(-\Lambda) + F(0) \right)
        \nonumber
        \\
        &\underset{g \to 1^+}{=}
         \frac{1}{2 \pi} \left( g - 1\right)^{-1/2}  + \cdots
        \,,
    \end{align}
    where $(\cdots)$ denotes non-singular terms.
\end{itemize}
\subsection{Spin Correlation Functions}\label{app:spin_correlation_functions}
The mapping to free fermions allows us to study equal-time spin correlation functions.  Of particular interest are $\sigma^z$ correlation functions within the ground state of the cTFIM, which in the fermionic language take the form
\begin{equation}\label{eqn:app_spin_spin_correlation_function_string}
C(i, j) = 
    \langle \sigma^z_i \sigma^z_j
    \rangle = 
    \left(
    \ket{0}_H\right)^\dagger \gamma_i \prod_{k=i}^{j-1} i \eta_k \gamma_k \gamma_j \ket{0}_H \,.
\end{equation}
Here $\ket{0}_H$ is the fermionic vacuum~\eqref{eqn:app_non_exceptional_point_vacuum_solution}. Furthermore, it will be useful to set the normalization such that $\left(\ket{0}_H\right)^\dagger \ket{0}_H = 1$.

The correlation function~\eqref{eqn:app_spin_spin_correlation_function_string} takes the form of a string operator in the fermionic language, which can be evaluated in the free theory through Wick's Theorem.  One finds that~\cite{mccoy_pfaffian}
\begin{equation}
    \langle \sigma^z_i \sigma^z_j \rangle = 
    \frac{1}{4}
    \mathrm{Pf} K^{(ij)}
    = \frac{1}{4} \sqrt{\det K^{(ij)}}
    \,,
\end{equation}
where $K^{(ij)}$ is a $2(j-i)$-by-$2(j-i)$ anti-symmetric matrix consisting of elementary $2$-pt correlators with the block structure
\begin{equation}\label{eqn:app_correlation_matrix}
    K^{kij)} = 
    \begin{pmatrix}
        S & M \\
        - M^T & Q
    \end{pmatrix}
    \,,
\end{equation}
where the $(j-i)$-by-$(j-i)$ matrices $S, M, Q$ are defined in terms of functions $F$ and $G$
\begin{align}
        S_{a, b} &= 
        \delta_{a, b} - \langle \eta_a \eta_b \rangle 
        = 2 i \Im F(a-b)
        \\
        Q_{a, b} &= 
        \langle \gamma_a \gamma_b \rangle - \delta_{a, b}
        = 2 i \Im F(a-b)
        \\
        M_{a, b} &= 
        i \langle \gamma_{b+1} \eta_a \rangle
        \\
        &=
        \delta_{a, b+1} - 2 \left[
        \Re F(a - b - 1) + G(a - b - 1)
        \right]
        \nonumber
    \,,
\end{align}
where
\begin{align}\label{eqn:app_F_and_G_functions}
F(r) &= \langle c_r c_0\rangle
= 
-
\frac{1}{L}
\sum_k
\frac{\sin k \sin k r(ig + \cos k - (\epsilon^+/ 2)^*)}
{\sin^2 k + | i g - \cos k + \epsilon^+/2|^2}
\nonumber
   \\ 
   G(r) &= \langle c_r c^\dagger_0 \rangle
   =
\frac{1}{L}
\sum_k
\frac{\cos k r
| i g - \cos k + \epsilon^+/2|^2
}
{\sin^2 k + 
| i g - \cos k + \epsilon^+/2|^2
}
\,.
\end{align}

\begin{figure*}[t]
\centering
\includegraphics[width=1 \textwidth  ]{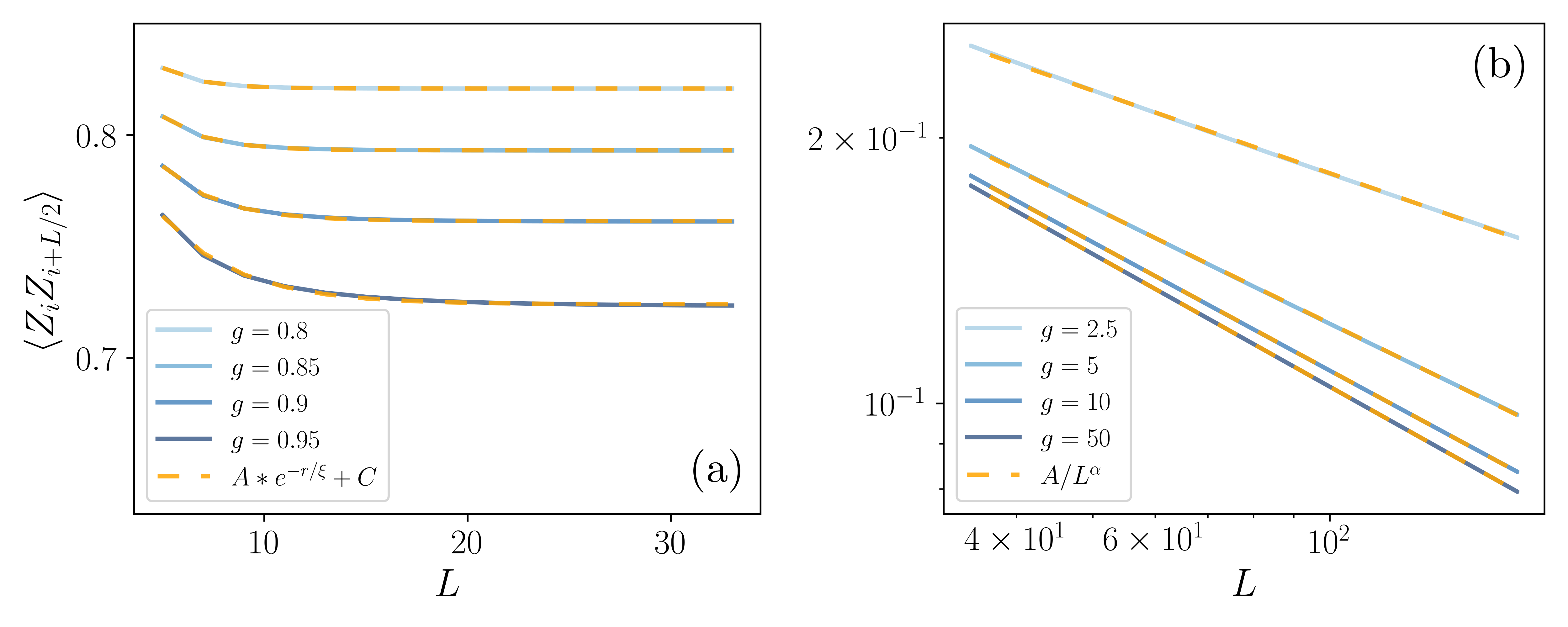}
    \caption{
    Spin-spin correlator $C(i, j)$ at half system size as a function of $L$ for different field strength $g$.
    In (a), we plot for $g$ within the ferromagnetic phase and fit to $A e^{-r/\xi} + C$ demonstrating long-range order.
    In (b), we plot for $g$ within the paramagnetic phase and fit to $A / L^\alpha$ demonstrating a quasi-long range ordered phase with $\alpha$ varying as a function of $g$.
    }\label{fig:spin_correlator_representative_behavior}
\end{figure*}

\begin{figure*}[t]
\centering
\includegraphics[width=1 \textwidth  ]{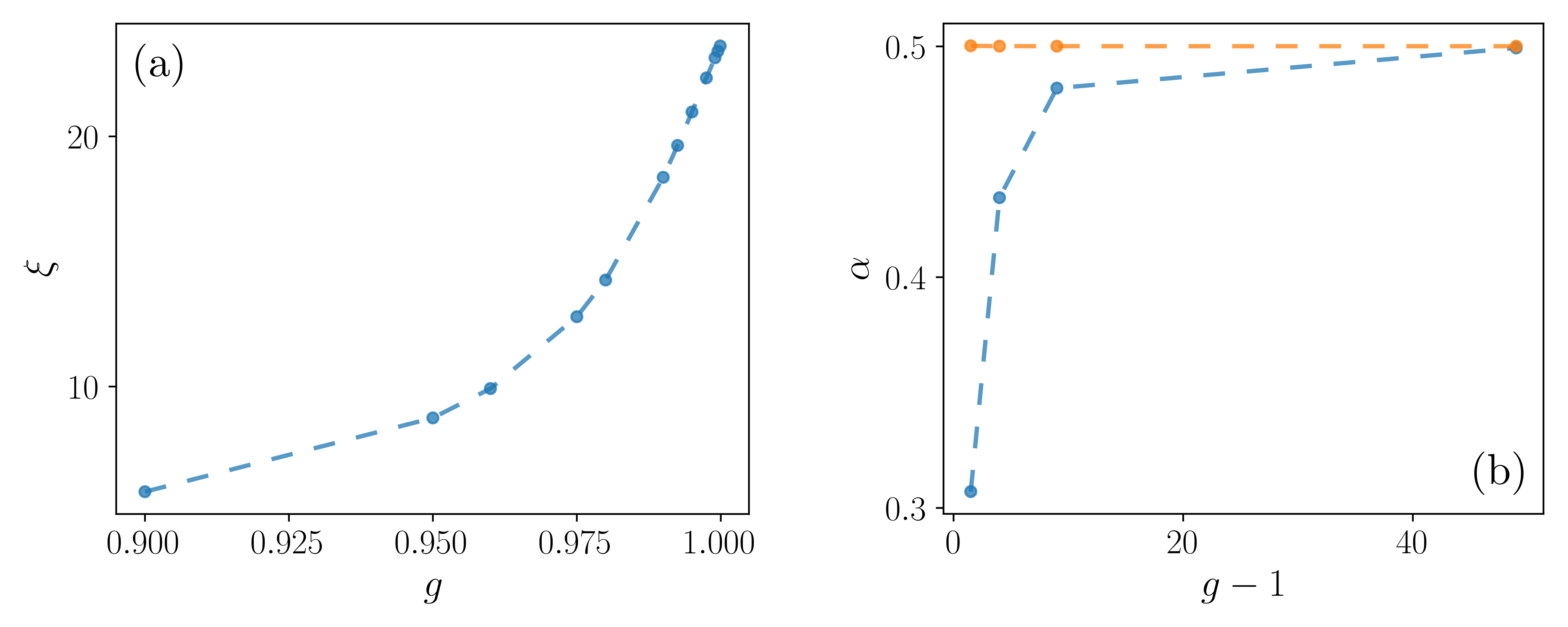}
    \caption{
    In (a), we extract the correlation length within the ferromagnetic phase, and show that it approaches a non-diverging constant as the transition is approached.  In (b), we extract the powerlaw exponent of the correlation function $C(i, j)$ (blue) and of the bicorrelator $B(i, j)$ (orange) in the paramagnetic phase.
    }
    \label{fig:spin_correlators_fit_trends}
\end{figure*}

\begin{figure*}[t]
\centering
\includegraphics[width=1 \textwidth  ]{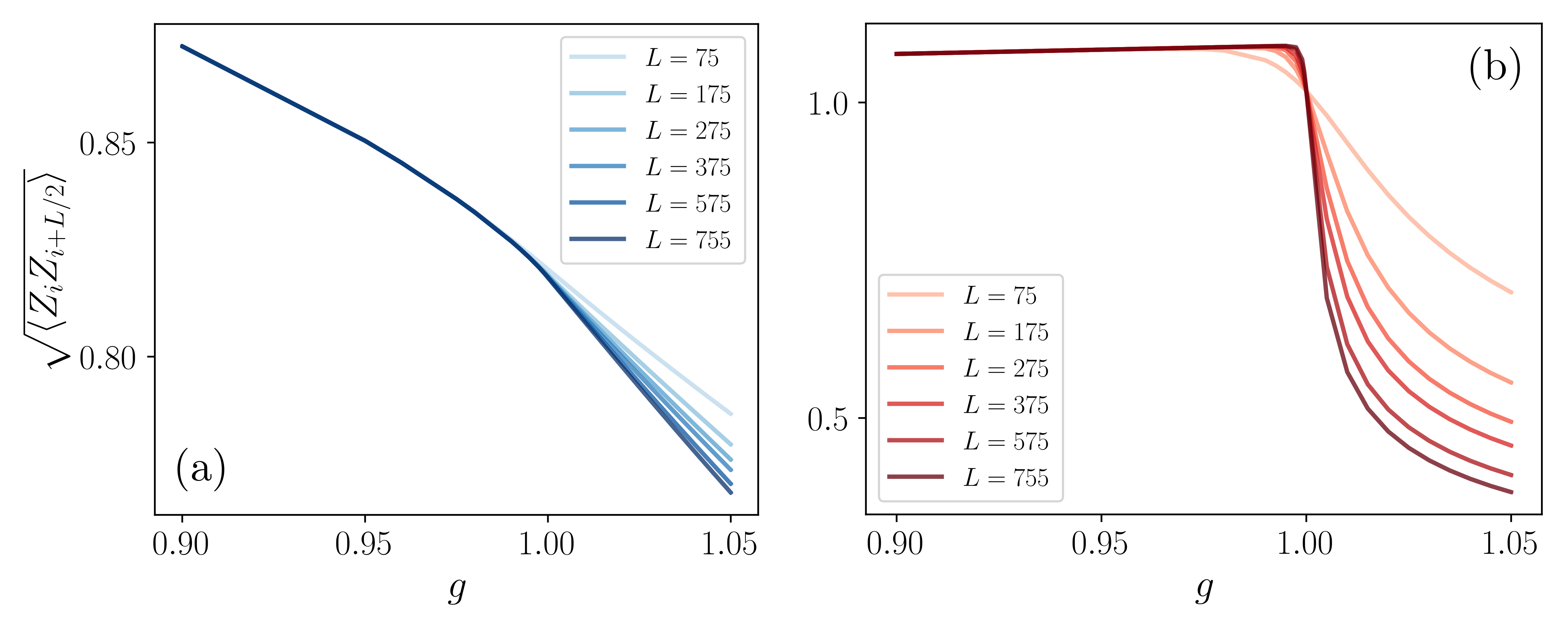}
    \caption{
    We plot the magnetization $m$, estimated by $\sqrt{\langle \sigma^z_0 \sigma^z_{L/2}\rangle}$ at large, but finite $L$, for various $g$ near the transition.
    In (a), $\langle \sigma^z_0 \sigma^z_{L/2} \rangle$ is defined by $C(i, j)$ while in (b) it is defined with $B(i, j)$.
    While $m$ clearly appears to decay for $g>1$, it appears to saturate to a non-zero constant for $g<1$ even as $g \to 1^-$, consistent with the interpretation of a first-order transition from this direction.
    }
    \label{fig:magnetization_versus_g}
\end{figure*}

In Figure~\ref{fig:spin_correlator_representative_behavior}, we plot the spin-spin correlation function versus distance for different values of $g$, demonstrating an exponential decay towards a constant in the ferromagnetic phase and a powerlaw decay towards zero in the paramagnetic phase.
In Figure~\ref{fig:spin_correlators_fit_trends}, we extract the correlation length within the ferromagnetic phase and the powerlaw exponent in the paramagnetic phase, plotted as a function of $g$.
Finally, in Figure~\ref{fig:magnetization_versus_g}, we plot the magnetization, defined by $\lim_{L \to \infty} \sqrt{\langle \sigma^z_{i} \sigma^z_{i+L/2} \rangle}$, and numerically estimated at finite $L$, for varying $g$.

On the other hand, we may also compute the ``bicorrelator" $B(i, j)$ defined by replacing $\left(\ket{0}_H\right)^\dagger$ with the left-ground eigenstate $\bra{0}_H$ in~\eqref{eqn:app_spin_spin_correlation_function_string}, with the normalization now fixed by $\bra{0}_H \ket{0}_H = 1$.
The bicorrelator can again be evaluated by repeated Wick contractions, however, one finds this time that the $S$ and $Q$ submatrices in~\eqref{eqn:app_correlation_matrix} vanish.
Therefore, one finds that $B(i, j) = \det M$, where $M$ is the $(j-i)$-by-$(j-i)$ matrix given by
\begin{align}
    M_{a, b} 
    &= i \bra{0}_H \gamma_{b+1} \eta_{a} \ket{0}_H 
    \\
    &= \frac{2}{L}
    \sum_k
    \frac{\sin k \sin kr  + \cos k \cos k r - i g \cos k r}{\epsilon^+}
    \nonumber
    \,,
\end{align}
where $r = a - b - 1$.
We numerically evaluate the bicorrelator, and find similar behavior as the ordinary correlator, with the following distinctions.
Namely, in the ferromagnetic phase, the correlator grows as a function of distance before saturating to a constant magnetization, which increases as a function of increasing field strength before dropping discontinuously to zero at the transition.
Additionally, the paramagnetic phase is again characterized with powerlaw correlators, however the exponent is approximately $\frac{1}{2}$ and numerically appears to be insensitive of the field strength.
This behavior is plotted in Fig.~\ref{fig:spin_correlators_fit_trends} and Fig.~\ref{fig:magnetization_versus_g}.

\section{Perturbations of the One-Dimensional Model}\label{app:perturbations}
While the specific dynamics considered in Sec.~\ref{sec:one_dim} admits a clean mapping into the cTFIM~\eqref{eqn:non_hermitian_hamiltonian}, additional perturbations can appear.  
We consider the simple case where the perturbations preserve the weak $\mathbb{Z}_2$ symmetries and the global $\mathbb{Z}_2$ Ising symmetry of the basic model.
The former is required for our mapping of the unconditional dynamics into the non-Hermitian Hamiltonian~\eqref{eqn:non_hermitian_hamiltonian} while the latter is used to formulate the over-to-underdamped transition as an ordinary ordering transition (see Sec.~\ref{sec:spin_spin_correlations_1d}).
Even in this restricted setting, different classes of perturbations can be added, which we discuss in turn.

First, generic symmetry-preserving perturbations of the dynamics map to symmetry-preserving interaction terms in the non-Hermitian fermionic Hamiltonian obtained after fixing the weak-symmetry $\mu^x_j = -1$ and the Jordan-Wigner transform.
Perturbations of this type include non-local or multi-site dephasing gates.
An example of the former is the dephasing of $\sigma^z_i \sigma^z_j$ while the latter includes the dephasing of $\sigma^z_i \sigma^z_{i+1} \sigma^z_{i+2} \sigma^z_{i+3}$.
These perturbations correspond to adding the terms $\tau^z_i \tau^z_j$ and $\tau^z_i \tau^z_{i+1} \tau^z_{i+2} \tau^z_{i+3}$ to~\eqref{eqn:non_hermitian_hamiltonian}, respectively.
These terms correspond to adding interactions in the fermionic description~\eqref{eqn:general_post_jordan_wigner}.
The fate of the universality we observe under such perturbations is not clear due to absence of a well-behaved field theory (see Sec.~\ref{sec:field_theory}).

On the other hand, certain fine-tuned perturbations preserve the integrability of the model in addition to the $\mathbb{Z}_2$ symmetries.
For instance, one can add to the Lindbladian a term arising from jump operators proportional to $\sigma^z_j e^{i \pi \sigma^x_{j+1}/4} \sigma^z_{j+2}$.
Up to constant terms, this adds to the Lindbladian
\begin{align}
  \sum_j&  \Bigg(  \sigma^z_j \sigma^z_{j+2} \otimes \sigma^z_j \sigma^z_{j+2}
  +  \sigma^z_j \sigma^x_{j+1} \sigma^z_{j+2}
  \otimes 
  \sigma^z_j \sigma^x_{j+1} \sigma^z_{j+2}
  \nonumber
  \\
  &+ i
   \Big[\sigma^z_j \sigma^x_{j+1} \sigma^z_{j+2} \otimes \sigma^z_j \sigma^z_{j+2}  
   - \sigma^z_j \sigma^z_{j+2} \otimes  \sigma^z_j \sigma^x_{j+1} \sigma^z_{j+2} 
   \Big]
   \Bigg)
   \nonumber
   \\
   &= \sum_j i \tau^z_j \tau^x_{j+1} \tau^z_{j+2}
   \,,
\end{align}
where in the second line we have fixed $\mu^x_j = -1$.
The effective non-Hermitian Hamiltonian in the fermionic representation~\eqref{eqn:app_free_fermion_real_space} is modified and takes the form
\begin{equation}
    H(g) = i \sum_i \eta_i \gamma_{i+1} +  g \sum_i \gamma_i \eta_i + \phi \sum_j  \eta_j \gamma_{j+2} \,,
\end{equation}
where $\phi$ parameterizes the strength of the perturbation.
Since the perturbation is integrable, it can be studied exactly in the way described in App.~\ref{app:free_fermion} and App.~\ref{app:long_time_limit}.
We omit the detailed manipulations here, but note that both the diverging decay rate at large-$T$, as well as the spin-spin correlators exhibit the same behaviors as in the unperturbed model.

In the case of the decay rate, its value at large-$T$ is determined by the real part of the ground state energy~\eqref{eqn:complex_energy_spectrum} modified by the addition of higher harmonics in the presence of generic integrable perturbations.
As a result, both the value of $g_c(\phi)$, and the point in momentum space $k_\mathrm{EXP}(\phi)$ at which the exceptional point occurs, shifts.
The analysis of App.~\ref{app:decay_rate_singularity} follows since the singularity is determined only by the behavior of the spectrum in the neighborhood of $k_\mathrm{EXP}(\phi)$ as $g$ is sent to $g_c(\phi)$ from above.

For the spin-spin correlators, the functions in Eq.~\eqref{eqn:app_F_and_G_functions} are also modified by the presence of higher harmonics.
Numerically studying the resulting correlators, we observe qualitatively similar behavior as in the unperturbed model.
For instance, the transition is still between a ferromagnetic phase with long-range order into a paramagnetic phase with quasi-long range order with exponent varying with $g$.

Finally, certain perturbations of the dynamics are trivialized after fixing the weak-symmetry.
These include dephasing in the $X$-basis or nearest-neighbor dephasing of $\sigma^y_j \sigma^y_{j+1}$ or $\sigma^x_j \sigma^y_{j+1}$.
For example, single-site dephasing in the $X$-basis adds a term proportional to $\mu^x_j$ in the Lindbladian, which becomes a constant after fixing $\mu^x_j = -1$.
Similarly, dephasing of $\sigma^y_j \sigma^y_{j+1}$ adds the term $\mu^x_j \mu^x_{j+1} \tau^z_j \tau^z_{j+1}$ whose only role is to renormalize the strength of the Ising coupling in~\eqref{eqn:non_hermitian_hamiltonian} after fixing the weak-symmetry.
\onecolumngrid
\section{Derivation of Field Theory}\label{app:mean_field}
\subsection{Derivation of Effective Action}\label{app:effective_action}
Expanding the action~\eqref{eqn:euclidean_action} for small fluctuations $\varphi(\tau) = \phi_0 + \phi(\tau)$ around the saddlepoint, we have the effective action to quadratic order
\begin{equation}
    S_\mathrm{eff}[\phi] = S[\phi_0] + \frac{1}{2}
    \int d \tau d \tau' \sum_{ij} 
    \frac{\delta^2 S}
    {
    \delta \varphi_i(\tau) \delta \varphi_j(\tau')
    }
    \Bigg|_{\varphi = \phi_0}
    \phi_i(\tau)
    \phi_j(\tau')
    + \mathcal{O}\left(\delta^3 \right)
    \,,
\end{equation}
using the saddlepoint condition $\delta S = 0$.
The second variation can be computed 
\begin{equation}
    \frac{\delta^2 S}
    {
    \delta \varphi_i(\tau) \delta \varphi_j(\tau')
    }
    \Bigg|_{\varphi = \phi_0}
   =
   J_{ij}^{-1} \delta(\tau - \tau')
   - 
   \langle \tau^z_i(\tau) \tau^z_i(\tau') \rangle^c_h
   \delta_{ij}
   \,,
\end{equation}
where 
$
   \langle \tau^z_i(\tau) \tau^z_i(\tau') \rangle^c_h
$
is the connected correlation function with respect to the mean-field Hamiltonian $h$~\eqref{eqn:local_mean_field_hamiltonian}.
Defining $C(\omega, \nu)$ as the connected correlator in Fourier space, as well as the Fourier transform of the nearest-neighbor Ising coupling $J(\vec{q}) = 2 \sum_{\alpha=1}^{d} \cos q_\alpha$, we obtain the effective action in momentum space
\begin{equation}
    S_\mathrm{eff} = 
    \frac{1}{2} 
    \int
    \frac{d^d q d \omega d \nu}{(2 \pi)^{d+2}}
    \phi(q, \omega)
    \left[
    \frac{2 \pi \delta(\omega + \nu)}{J(\vec{q})} - C(\omega, \nu)
    \right]
    \phi(-q, \nu)
    \,,
\end{equation}
which is the form quoted in the main text.
\subsection{Correlators in Local Mean-Field Hamiltonian}\label{app:mean_field_correlators}
It will be useful to compute correlation functions of $\tau^z(t)$ with respect to the mean-field Hamiltonian
\begin{equation}
    -h[\phi_0] = 
    \phi_0 \tau^z + i g \tau^x \,.
\end{equation}
Away from the exceptional point $g = \phi_0$, the non-Hermitian Hamiltonian has two independent right-eigenvectors
\begin{equation}
    \ket{\pm} = 
    \begin{pmatrix}
        1 \\
        \frac{i}{g} 
        \left[
        \phi_0
        \mp \sqrt{\phi_0^2 - g^2}
        \right]
    \end{pmatrix}
    \,,
\end{equation}
with corresponding eigenvalues $E_\pm = \pm \sqrt{\phi_0^2 - g^2}$.
We can also find left-eigenvectors
\begin{equation}
    \bra{\pm} = 
    \frac{i g}{2 \sqrt{\phi_0^2 - g^2}}
    \begin{pmatrix}
        - \frac{i}{g}
        \left[
        \phi_0
        \pm \sqrt{\phi_0^2 - g^2}
        \right]
        & 1
    \end{pmatrix}
    \,,
\end{equation}
which are normalized such that $\braket{\pm}{\mp} = \braket{\mp}{\pm} = 0$ and $\braket{\pm}{\pm} = 1$.
\begin{itemize}
    \item \textbf{Partition Function--}The partition function $Z = \Tr \exp - \int d \tau h(\tau)$ is given by
    \begin{equation}
        Z = \sum_\pm
        \bra{\pm}
        e^{- T h}
        \ket{\pm}
        = 2 \cosh 
        T \sqrt{\phi_0^2 - g^2}
        \,.
    \end{equation}
    \item \textbf{Spin Expectation Value--}In order to evaluate the spin expectation value from the insertion of $\tau^z$ into the trace, we use the identity
    \begin{equation}
        \tau^z \ket{\pm} = 
        \pm 
        \frac{\phi_0}{\sqrt{\phi_0^2 - g^2} }
        \ket{\pm}
        +
        \left(
        1  \mp
        \frac{\phi_0}{\sqrt{\phi_0^2 - g^2} }
        \right)
        \ket{\mp}
        \,.
    \end{equation}
    Using the cyclicity of the trace, this implies
    \begin{equation}
    \begin{split}
        \Tr \tau^z(T)
        e^{- T h}
        &=
        \sum_\pm
        \bra{\pm}
        \tau^z
        \ket{\pm}
        e^{T E_\pm}
        = \sum_\pm
        \pm
        \frac{\phi_0}{\sqrt{\phi_0^2 - g^2} }
        e^{\pm T \sqrt{\phi_0^2 - g^2}}
        \\
        &= 2 
        \frac{\phi_0}{\sqrt{\phi_0^2 - g^2} }
        \sinh 
        T \sqrt{\phi_0^2 - g^2}
    \end{split}
    \,.
    \end{equation}
    This implies that
    \begin{equation}
        \langle \tau^z(t) \rangle_{h}
        =
        \frac{\phi_0}{\sqrt{\phi_0^2 - g^2}} 
        \tanh 
        T \sqrt{\phi_0^2 - g^2}
        \,.
    \end{equation}
    \item \textbf{Spin-Spin Correlation Function--}The spin-spin correlation function is obtained by inserting $\tau^z(t_2)$ and $\tau^z(t_1)$ within the trace.
    Defining $\Delta = t_2 - t_1 > 0$ and again using the cyclicity of the trace, 
    \begin{align}
        \Tr \tau^z(T)
        e^{- \Delta h}
        \tau^z(T- \Delta)
        e^{- (T - \Delta) h}
        &= 
        \sum_\pm
        \bra{\pm}
        \tau^z
        e^{-\Delta h}
        \tau^z
        \ket{\pm}
        e^{(T- \Delta)E_\pm}
        \nonumber
        \\
        &=
        \sum_\pm
        \bra{\pm}
        \tau^z
        \ket{\pm}
        \left( \pm 
        \frac{\phi_0}{\sqrt{\phi_0^2 - g^2} }
        \right)
        e^{T E_\pm}
        +
        \bra{\pm}
        \tau^z
        \ket{\mp}
        \left( 1 \mp
        \frac{\phi_0}{\sqrt{\phi_0^2 - g^2}} 
        \right)
        e^{(T- 2\Delta)E_\pm}
        \nonumber
        \\
        &= 
        \sum_\pm
        \frac{\phi_0^2}{\phi_0^2 - g^2} 
        e^{T E_\pm}
        -
        \frac{g^2}{\phi_0^2 - g^2} 
        e^{(T- 2\Delta)E_\pm}
        \nonumber
        \\
        &=
        2 
        \frac{\phi_0^2}{\phi_0^2 - g^2} 
        \cosh T \sqrt{\phi_0^2 - g^2}
        -
        2
        \frac{g^2}{\phi_0^2 - g^2} 
        \cosh (T - 2 \Delta) 
        \sqrt{\phi_0^2 - g^2}
    \,.
    \end{align}
    Therefore, we conclude that
    \begin{equation}
        \langle \tau^z(t_2) \tau^z(t_1) \rangle_h = 
        \frac{\phi_0^2}{\phi_0^2 - g^2} 
        -
        \frac{g^2}{\phi_0^2 - g^2} 
        \left[
        \cosh 2 \Delta
        \sqrt{\phi_0^2 - g^2}
        - 
        \tanh T
        \sqrt{\phi_0^2 - g^2}
        \sinh 2 \Delta 
        \sqrt{\phi_0^2 - g^2}
        \right]
        \,.
    \end{equation}
    On the other hand, the connected correlator is obtained by subtracting off $\langle \tau^z \rangle_h^2$ and has the form
    \begin{equation}\label{eqn:app_connected_correlator_mf_hamiltonian}
    \begin{split}
        \langle \tau^z(t_2) \tau^z(t_1) \rangle_h^c &= 
        \frac{\phi_0^2}{\phi_0^2 - g^2} 
        \frac{1}{\cosh^2 T \sqrt{\phi_0^2 - g^2}}
        \\
        &-
        \frac{g^2}{\phi_0^2 - g^2} 
        \left[
        \cosh 2 \Delta
        \sqrt{\phi_0^2 - g^2}
        - 
        \tanh T
        \sqrt{\phi_0^2 - g^2}
        \sinh 2 \Delta 
        \sqrt{\phi_0^2 - g^2}
        \right]
    \end{split}
        \,.
    \end{equation}
\end{itemize}
\subsection{Solving for Saddlepoints and Stability}\label{app:saddlepoints_and_stability}
As discussed in the main text, there are three types of saddlepoints.
\begin{itemize}
    \item \textbf{Class A--}We solve the saddlepoint condition~\eqref{eqn:saddle_point} under the assumption that $\phi_0 > g$.
    In this case, the saddlepoint condition
    \begin{equation}
        \frac{1}{2d} 
        \sqrt{\phi_0^2 - g^2}
        =
        \tanh T
        \sqrt{\phi_0^2 - g^2}
        \,,
    \end{equation}
    has a well-defined $T \to \infty$ limit, where we find that
    \begin{equation}
        \phi_0 = \sqrt{g^2 + 4d^2} \,.
    \end{equation}
    The action has the value
    \begin{equation}
        S[\phi_0] = 
        \frac{1}{2}
        \phi_0^2
        T
        \sum_{ij}
        J^{-1}_{ij}
        - N \log
        \left(
        2 \cosh 
        T
        \sqrt{\phi_0^2 - g^2} 
        \right)
        \,,
    \end{equation}
    which gives for $T \to \infty$
    \begin{equation}
        S[\phi_0] = 
        \frac{\phi_0^2}{2} T \frac{N}{2d} - N T \sqrt{\phi_0^2 -g^2} = 
        T N d \left[
        \left(
        \frac{g}{2d}
        \right)^2
        -1
        \right]
        \,.
    \end{equation}
    at the saddlepoint.
    This is positive when $g > 2d$ but negative for $g < 2d$.
    To study the stability, we consider the
    connected correlator~\eqref{eqn:app_connected_correlator_mf_hamiltonian},
    which has in the $T \to \infty$ limit the form
    \begin{equation}
        C(\tau, \tau') = \frac{g^2}{g^2 - \phi_0^2} e^{- 2 \sqrt{\phi_0^2 - g^2} |\tau - \tau'|}
        \implies
        C(\omega, \nu) = - \frac{2 g^2}{d} 
        \frac{
        \delta(\omega + \nu)
        }
        {
        \omega^2 + (4d)^2
        }
        \,.
    \end{equation}
    This leads to the effective action
    \begin{equation}
         S_\mathrm{eff} 
         = \frac{1}{2}
         \int
         \frac{
         d^d q d \omega
         }
         {
         (2 \pi)^{d+1}
         }
         | \phi(q, \omega) |^2
         \left(
         \frac{1}{2d} + \frac{q^2}{4 d^2} + \frac{2g^2}{4 d}
         \frac{1}{\omega^2 + (4d)^2}
         \right)
         \,,
    \end{equation}
    where we've expanded $J(\vec{q})$ for small $\vec{q}$, but have not expanded the third term to make manifest that $S_\mathrm{eff}$ is a positive function of $\omega$.
    \item \textbf{Class B--}One can also see that the saddlepoint condition is satisfied by $\phi_0 = 0$.
    The action evaluated at this saddle gives
    \begin{equation}
        S[\phi_0 = 0] = - N \log \left(2 \cos g T\right)
        \,,
    \end{equation}
    which could in general, be complex.
    If we focus on the real part, we see that it is bounded from below by $\Re S[\phi_0 = 0] > - N \log 2$.

    On the other hand, the connected correlator at this saddle takes the form
    \begin{equation}
        C(\tau, \tau') = 
        \cos \left(2 g (\tau - \tau') \right)
        +
        \tan \left( g T \right)
        \sin \left(2 g (\tau - \tau') \right)
        \,.
    \end{equation}
    The Fourier transform of cosine is symmetric under $\omega \to - \omega$ while the Fourier transform of sine is anti-symmetric.
    On the other hand, the rest of the terms in the momentum-space effective action are symmetric under $\omega \to - \omega$ implying that only the cosine contribution survives.
    Therefore, we find the effective action, again for $q$ small
    \begin{equation}
         S_\mathrm{eff} 
         = \frac{1}{2}
         \int
         \frac{
         d^d q d \omega
         }
         {
         (2 \pi)^{d+1}
         }
         | \phi(q, \omega) |^2
         \left(
         \frac{1}{2d} + \frac{q^2}{4 d^2} 
         - \frac{1}{2}
         \left[
         \delta\left(\omega + \frac{g}{\pi}\right)
         +
         \delta\left(\omega - \frac{g}{\pi}\right)
         \right]
         \right)
         \,.
    \end{equation}
    Note that for $d > 1$, this implies that the saddle is unstable to fluctuations of frequency $\omega = \pm g / \pi$.
    \item \textbf{Class C--}Finally, we solve the saddlepoint condition assuming that $\phi_0 < g$.
    This results in the mean-field condition
    \begin{equation}
        \frac{1}{2d}
        \sqrt{g^2 - \phi_0^2}
        =
        \tan
        T
        \sqrt{g^2 - \phi_0^2}
        \,,
    \end{equation}
    which has an infinite number of solutions satisfying $\phi_0 < g$ when $T$ is large.
    At this saddle, the action is
    \begin{equation}
        S[\phi_0] = 
        \frac{1}{2} T \frac{N}{2d} \phi_0^2
        - N \log \frac{2}{\sqrt{1 + \tan^2 T \sqrt{g^2 - \phi_0^2}}}
        =
        \frac{NT \phi_0^2}{4d}
        - N \log
        \frac{
        4d
        }
        {
        \sqrt{4d^2 + g^2 - \phi_0^2}
        }
        \,,
    \end{equation}
    which again may be complex, but examining the real part tells us that it is exponentially suppressed at large $T$.

    On the other hand, the connected correlator has the form
    \begin{equation}
        - \frac{\phi_0^2}{g^2 - \phi_0^2}
        \left(
        1 + \frac{g^2 - \phi_0^2}{4d^2}
        \right)
        + \frac{g^2}{g^2 - \phi_0^2}
        \left[
        \cos 2 (\tau - \tau') \sqrt{g^2 - \phi_0^2}
        +
        \frac{\sqrt{g^2 - \phi_0^2}}{2d}
        \sin 2 (\tau - \tau') \sqrt{g^2 - \phi_0^2}
        \right]
        \,.
    \end{equation}
    This gives us the effective action for $q$ small
    \begin{equation}
    \begin{split}
         S_\mathrm{eff} 
         = \frac{1}{2}
         \int
         \frac{
         d^d q d \omega
         }
         {
         (2 \pi)^{d+1}
         }
         | \phi(q, \omega) |^2
         &
         \Bigg(
         \frac{1}{2d} + \frac{q^2}{4 d^2} 
        + \frac{\phi_0^2}{g^2 - \phi_0^2}
        \left(
        1 + \frac{g^2 - \phi_0^2}{4d^2}
        \right)
        \delta(\omega)
        \\
         &- \frac{1}{2}
         \frac{g^2}{g^2 - \phi_0^2}
         \left[
         \delta\left(\omega + \frac{\sqrt{g^2 - \phi_0^2}}{\pi}\right)
         +
         \delta\left(\omega - \frac{\sqrt{g^2 - \phi_0^2}}{\pi}\right)
         \right]
         \Bigg)
    \end{split}
         \,.
    \end{equation}
    Since $\frac{g^2}{g^2 - \phi_0^2} > 1$, then we conclude that for $d > 1$, the saddle is unstable to fluctuations of frequency $\omega = \pm \sqrt{g^2 - \phi_0^2} / \pi$.
\end{itemize}
\twocolumngrid
\bibliography{biblio}

\end{document}